\documentclass[oneocolumn,preprintnumbers,amsmath,amssymb,nofootinbib]{revtex4}
\pdfoutput=1

\usepackage{amsmath,latexsym,amssymb,amsfonts}
\usepackage{graphicx,color}
\usepackage{bm}
\usepackage{gensymb}

\begin{document}

\title{\textbf{Matching Radial Geodesics in Two Schwarzschild Spacetimes (e.g. Black-to-White Hole Transition) or Schwarzschild and de Sitter Spacetimes (e.g. Interior of a Non-singular Black Hole)}}
\author{
Wei-Chen Lin$^{a,b}$\footnote{{\tt archennlin@gmail.com}}
and
Dong-han Yeom$^{a,c,d}$\footnote{{\tt innocent.yeom@gmail.com}}
}
\affiliation{
$^{a}$\footnotesize{Center for Cosmological Constant Problem, Pusan National University, Busan 46241, Republic of Korea}\\
$^{b}$\footnotesize{Department of Physics, Pusan National University, Busan 46241, Republic of Korea}\\
$^{c}$\footnotesize{Department of Physics Education, Pusan National University, Busan 46241, Republic of Korea}\\
$^{d}$\footnotesize{Research Center for Dielectric and Advanced Matter Physics, Pusan National University, Busan 46241, Republic of Korea} 
}

\begin{abstract}
In this article, we study the trajectory equations of the bounded radial geodesics in the generalized black-to-white hole bounce with mass difference and the Schwarzschild-to-de Sitter transition approximated by the thin-shell formalism.  We first review the trajectory equations of the general radial geodesics in Kruskal-Szekeres (like) coordinates of the Schwarzschild and de Sitter spacetimes, respectively. We then demonstrate how one relates the radial geodesics on each side of the shell by correctly choosing the constants of integration after performing the two coordinate transformations mentioned in our previous work \cite{Lin:2023ztq}. We next show that the coordinate system used in the resulting Penrose diagram has no illness at the thin shell but instead creates a degeneracy between the timelike geodesics and null geodesics at the event horizons where the second transformation is applied. Due to this problem, we conclude that a global conformal coordinate chart for the spacetime connected via a static spacelike thin shell in general does not exist, except for some special cases. Since this is an extension of our work in Ref.~\cite{Lin:2023ztq}, we focus on Schwarzschild and de Sitter spacetimes, though the method should be applicable to any cut-and-pasted spacetime connected with a static spacelike thin shell.
 
\end{abstract}

\maketitle

\newpage

\tableofcontents

\section{Introduction}\label{Sec:Intro}

The thin-shell approximation is a very useful tool to describe the dynamic interactions between gravity and matter fields. The fundamental formalism was developed by Israel \cite{Israel:1966rt}. This technique has many applications. For example, thin shells can describe the dynamics of vacuum bubbles \cite{Blau:1986cw} as well as their tunneling processes \cite{Fischler:1990pk, Farhi:1989yr}. The tunneling of thin shells can be related to the issues of quantum gravity; for example, related to the Euclidean path integral approach \cite{Gregory:2013hja} and the black hole information loss paradox  \cite{Sasaki:2014spa,Chen:2022ric}. Not only timelike shells but also spacelike shells are useful to understand the quantum gravity-inspired internal structures of black holes \cite{Frolov:1988vj, Balbinot:1990zz}; especially, loop quantum gravity-inspired models might be well approximated by spacelike thin shells \cite{Brahma:2018cgr}. It might be possible to extend the thin-shell formalism to the Vaidya-de Sitter junction, which has interesting applications for dynamical evaporating black holes  \cite{Chen:2017pkl, Brahma:2019oal}.

Even though there is a lot of work related to thin shells as well as their gravitational dynamics, the precise and exact description of the coordinate transformations was less emphasized, because such a coordinate description is technically very difficult. Nevertheless, in Ref.~\cite{Lin:2023ztq}, Stojkovic and we found that in some special cases where a static spacelike shell connects two spacetimes, the coordinate transformations required to construct the corresponding Penrose diagrams can be made explicit and clear. By using the generalized black-to-white hole bounce and Schwarzschild-to-de Sitter transition as examples, we demonstrated the resulting Penrose diagrams are free from any explicit or implicit illness at the thin shell. 

As the companion work, here we discuss the construction of the trajectory equations of a bounded radial geodesic in the Penrose diagrams of the above-mentioned examples, and we show that the trajectory equation serves as a tool to examine the wellness of the resulting Penrose diagram. Together, the transformations and the trajectory equations of the timelike radial geodesics will be useful when we need to \textit{quantitatively} describe spacetime and geodesics, and interactions between geodesics.
For instance, the timelike radial geodesics and the coordinate systems adapted to them have been emphasized to be useful in some applications  \cite{Kraus:1994fh,Parikh:1999mf,Nielsen:2005af}. In these coordinate systems, the temporal coordinate is replaced by the proper time experienced by the free-falling observer moving along a family of timelike radial geodesics.  Extending this, one can further use these radial geodesics as the coordinate lines penetrating event horizons, which we will briefly review in Sec.~\ref{Sec: Bounded_Radial_Geodesics_Schwarzschild} and Appendix~\ref{Sec:Review}. It is an interesting question if these coordinates can be more suitable for the spacetimes constructed via a spacelike thin shell. In particular, the bounded timelike geodesics might be most useful in these scenarios since the resulting spacetimes are typically cyclic. Thus in principle, they are the geodesics penetrating every thin shell connecting the consecutive cycles and extending to both the infinite past and future. Then explicitly constructing the trajectory equations of the radial geodesics in these spacetimes should be the first step to analyze, what we do in the present work.

The work is organized as follows. In Sec.~\ref{Sec: Bounded_Radial_Geodesics_Schwarzschild}, we review some of the coordinate systems adapted to the free-falling observers in the Schwarzschild spacetime and utilize the method introduced by Martel and Poisson \cite{Martel:2000rn} to obtain the trajectory equations of the bounded timelike geodesic in the Kruskal diagram. Sec.~\ref{Sec:Bound_geodesics_matching} presents the construction of the bounded timelike radial geodesics in the generalized black-to-white hole bounce. We first quickly review the procedure introduced in Ref.~\cite{Lin:2023ztq} to create a Penrose diagram without illness at the transition surface. We then discuss how one constructs the trajectory equations of the bounded timelike radial geodesics in the resulting Penrose diagram. Firstly, by using the above-mentioned trajectory equations, we analytically show that the coordinates of the connected Penrose diagram introduced in Ref.~\cite{Lin:2023ztq} serve as a well-behaved coordinate chart covering the entire thin shell. We next show that those trajectories become either perpendicular or parallel at the event horizons influenced by the transformation fixing the implicit discontinuity at the thin shell introduced in Ref.~\cite{Lin:2023ztq}. In Sec.~\ref{Sec:Sch_to_dS}, we apply the similar analysis to the Schwarzschild-to-de Sitter transition and construct the trajectory equation of the bounded timelike radial geodesic in the corresponding Penrose diagram. Sec.~\ref{Sec:conclusion} is devoted to a discussion of our construction and the open directions it offers for the future. We work in the Planck unit: $G=c=\hbar=1$.


\section{The Timelike Bounded Radial Geodesics in the Null Kruskal-Szekeres Coordinates}\label{Sec: Bounded_Radial_Geodesics_Schwarzschild}

In this section, we first review the timelike radially infalling geodesics in the Schwarzschild spacetime to set up the convention and notations.  
We also briefly mention the coordinate systems adapted to the observers moving along those geodesics, \textit{i.e.}, the coordinate systems in which the temporal coordinate is the proper time $\tau$ measured by the free-falling observers. We next utilize the transformation derived by Martel and Poisson \cite{Martel:2000rn} on the bounded timelike infalling radial geodesics with $E<1$ to obtain the constant $\tau$ surfaces and the trajectories of the corresponding geodesics in the null Kruskal-Szekeres coordinates. We then perform the same construction for the outgoing type, and discuss the connection between those outgoing and infalling geodesics.

\subsection{Timelike radial geodesics in the Schwarzschild spacetime}

We start with the usual Schwarzschild metric:  
\begin{equation}\label{Schwarzschild_metric}
d s^2=-f(r)d t^2+f^{-1}(r)d r^2+r^2d\Omega^2, 
\end{equation}
where $f(r)=1-2M/r$ and $d\Omega^2=d\theta^2 +\sin^2{\theta}d\phi^2$. In this coordinate system, the four-velocity of an observer moving along with a radial timelike geodesic\footnote{We will neglect the term ``timelike'' for simplicity in the rest of the paper, since this is the only type of geodesics considered in this work.
} is given by
\begin{equation}\label{radial_geodesic_Deq}
\mathcal{U}^{\alpha}=\left(\frac{d t}{d \tau}, \frac{d r}{d \tau}, \frac{d \theta}{d \tau}, \frac{d \phi}{d \tau}\right)=\left(\frac{E}{f(r)}, -\epsilon \sqrt{E^2-f}, 0, 0\right),
\end{equation}
where $\tau$  and $E$ are the proper time and the conserved energy per unit mass\footnote{In the rest of this paper, we will simply call it the energy parameter related to a radial geodesic.} associated with the geodesic, respectively. Also notice that $\epsilon=1$ for the infalling geodesics and $\epsilon=-1$ for the outgoing geodesics.  For $E^2 \geq 1$, the conserved quantity (energy parameter) $E$ can be related to the observer's velocity at infinity $v_{\infty}$ as $E^2=1/(1-v^2_{\infty})$ by noticing that the Schwarzschild metric (\ref{Schwarzschild_metric}) reduces to the Minkowski metric at $r \rightarrow \infty$. For $E \leq 1$, $E$ is related to the maximal radius $R$ the geodesic can reach by the relation $E^2=1-2M/R$.  

We first notice that by integrating the differential equation from the $r$-component in Eq.~(\ref{radial_geodesic_Deq}), we obtain the following relation
\begin{equation}\label{tau_r_relation}
\tau -\tau_0= \epsilon \sqrt{\frac{R^3}{8M}}\left[\cos^{-1} \left(\frac{2r}{R}-1\right) + 2\sqrt{\frac{r}{R}-\frac{r^2}{R^2}} \; \right] \equiv \epsilon A(r),
\end{equation}
where $\tau_0$ is an integral constant and $E^2=1-2M/R$ has been used. On the other hand, the $t$-component cannot be solved directly. Instead, one can first rewrite the differential equations in Eq.~(\ref{radial_geodesic_Deq}) into 
\begin{equation}\label{t_r_Deq}
\frac{d t}{d r}=\frac{-\epsilon E}{f\sqrt{E^2-f}},
\end{equation}
so it reduces to a first order ordinary differential equation. Although one can solve this differential equation analytically already, for the purpose later, we follow the method used by Gautreau and Hoffmann \cite{Gautreau:1978zz} by combining $t$-component in Eq.~(\ref{radial_geodesic_Deq}) and Eq.~(\ref{t_r_Deq}) to have
\begin{equation}\label{constant_tau_surface_Deq}
\frac{d \tau}{d r}=E\frac{d t}{d r}+ \epsilon \frac{\sqrt{E^2-f}}{f}. 
\end{equation}
Solving this differential equation, one then obtains the equation of constant $\tau$ surface. This group of constant $\tau$ surfaces can serve as the new temporal coordinate, which has a natural interpretation as the proper time experienced by the infalling observers moving along with the family of geodesics specified by a given energy parameter $E$. The coordinate systems, in which the $t$-component used in the Schwarzschild metric is replaced by the proper time of a family of infalling geodesics, can be further categorized into three types. The simplest kind was introduced a century ago, the Painlev{\'e}-Gullstand coordinates \cite{1921CR....173..677P, Gullstrand:1922tfa}, which is the special case when $E=1$. Also, notice that by diagonalizing the Painlev{\'e}-Gullstand coordinates, one then obtains the  Lema{\^i}tre coordinates  \cite{1933ASSB...53...51L}.   
The second type is the Gautreau-Hoffmann coordinates \cite{Gautreau:1978zz}, which are constructed by geodesics with $E<1$.  Lastly, the third type is coordinates constructed by geodesics with $E>1$, which were studied and named the generalized Painlev{\'e}-Gullstand coordinates by Martel and Poisson \cite{Martel:2000rn}. A brief review of above mentioned coordinate systems constructed by the radially infalling geodesics with $E \leq 1$ is given in Appendix \ref{Sec:Review}.

\subsection{The radial infalling geodesics in the null Kruskal-Szekeres Coordinates}\label{SubSec:radial_infalling_in_Null_KS}

In  Ref.~\cite{Martel:2000rn}, Martel and Poisson derived the coordinate transformation between the generalized Painlev{\'e}-Gullstand coordinates and the null Kruskal-Szekeres coordinates. This transformation allows one to parametrically plot the constant $\tau$ surfaces in a Kruskal diagram. Here, we apply the same method on the bounded infalling radial geodesics with $E<1$, not only to obtain the constant $\tau$ surface of the Gautreau-Hoffmann type, but further to determine the trajectories of the corresponding geodesics in the null Kruskal-Szekeres coordinates.

\begin{figure}[h!]
	\centering
	\includegraphics[scale=0.4]{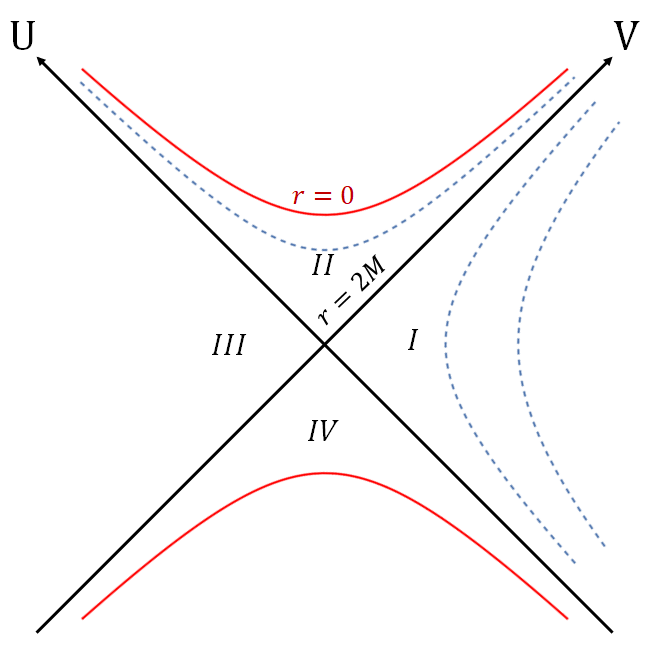}
	\caption{A Kruskal diagram. For infalling geodesics, the relevant regions are regions $I$ and $II$; for the outgoing geodesics, the relevant regions are regions $I$ and $IV$. The dotted lines are constant $r$ surfaces. }  
	\label{fig:Simple_KS}
\end{figure}

Firstly, the null Kruskal-Szekeres (KS) coordinates, $(V, U) $,
in terms of the Schwarzschild coordinates $(t,r)$ are given by \cite{Carroll:2004st}
\begin{equation}\label{Null_KS_in_r_t}
V=\pm e^{(t+r^{*})/4M};    \quad U=\mp e^{-(t-r^{*})/4M},
\end{equation}
where $r^{*}$ is the tortoise coordinate
\begin{equation}
r^{*} \equiv r + 2M \ln \left|\frac{r}{2M}-1\right|,
\end{equation}
and the plus/minus signs are directly determined by the region considered in a Kruskal diagram. See Fig.~\ref{fig:Simple_KS}. From Eq.~(\ref{Null_KS_in_r_t}), one can show that the null coordinates $(V, U) $ satisfy the relations
\begin{equation}\label{Kruskal_UV(r)}
UV=\left(1-\frac{r}{2M}\right)e^{r/2M},
\end{equation}
and 
\begin{equation}\label{Kruskal_U/V}
\frac{U}{V}=\mp e^{-t/2M},
\end{equation}
where the minus sign refers to the $r>2M$ region, and the plus sign refers to the $r<2M$ region.
Next, the metric of a maximally extended Schwarzschild solution in terms of the null KS coordinates $(V, U) $ is given by 
\begin{equation}\label{Kruskal_in_U_V}
d s^2=-\frac{16M^3}{r}e^{-r/2M}(d V d U+d U d V)+r^2 d \Omega^2,
\end{equation}
which is well-defined everywhere except only at the singularity. 

Now, we use the procedure from Ref. \cite{Martel:2000rn} to convert $(V, U)$ in Eq.~(\ref{Null_KS_in_r_t}) to functions of $\tau$ and $r$, where $\tau$ is the proper time measured by a group of radially free-falling observers with $E<1$. That is, we derive a mapping between the null KS coordinates and the Gautreau-Hoffmann coordinates. For a better comparison, we also adopt the parameter $p=1/E^2$ used in Ref. \cite{Martel:2000rn}. Using $p=1/E^2$, we then rewrite Eq.~(\ref{constant_tau_surface_Deq}) as
\begin{equation}\label{constant_tau_surface_Deq_in_P}
\sqrt{p}\frac{d \tau}{d r}=\frac{d t}{d r}+ \epsilon \frac{\sqrt{1-pf}}{f}. 
\end{equation}
By integrating this differential equation, the constant $\tau$ surface is given by the following relation
\begin{equation}\label{general_tau(t,r)}
\sqrt{p}\tau=t+ \epsilon \left(\Theta(r)+C \right), 
\end{equation}
with
\begin{equation}\label{Theta(r)_general}
\Theta(r) \equiv r\sqrt{1-pf}+\frac{2M(p-2)}{\sqrt{p-1}}\tan ^{-1}\left(\frac{\sqrt{1-pf}}{\sqrt{p-1}}\right)
+2M \ln{\left|\frac{1-\sqrt{1-pf}}{1+\sqrt{1-pf}}\right|}, 
\end{equation}
where $C$ is a constant of integration. Notice that the constant $\tau$ surface cannot be extended beyond the maximal radius $R$, since the factor $\sqrt{1-pf}$ becomes an imaginary number when $r>R$. Next, we decompose $\Theta(r)$ into 
\begin{equation}\label{general_tau(t,r)_in_S_r*}
\Theta(r)+C= S(r)+r^{*}.   
\end{equation}
Notice that $S(r)$ is finite at $r=2M$; hence, the divergence of $\Theta(r)$ at the horizon is solely due to the tortoise coordinate $r^*$.

For the timelike radially-infalling geodesics, we choose $\epsilon=1$ and use Eq.~(\ref{general_tau(t,r)_in_S_r*}) to rewrite Eq.~(\ref{general_tau(t,r)}) into
\begin{equation}\label{t+r^{*}}
t+r^{*}=\sqrt{p}\tau-S(r); \quad t-r^{*}=\sqrt{p}\tau-S(r)-2r^{*}. 
\end{equation}
By substituting the above relations into Eq.~(\ref{Null_KS_in_r_t}) for regions $I$ and $II$, we then have the KS coordinates in terms of functions of $(\tau, r)$ as
\begin{equation}\label{V_(tau, r)_BH}
V(\tau, r)=e^{\sqrt{p}\tau/4M}e^{-S/4M}
\end{equation}
and 
\begin{equation}\label{U_(tau, r)_BH}
U(\tau, r)=e^{r/2M}\left(1-\frac{r}{2M}\right)e^{-\sqrt{p}\tau/4M}e^{S/4M},
\end{equation}
which are well-behaved functions at the horizon since, as mentioned earlier, $S(r)$ is finite there. Then the trajectories of the constant $\tau$ surfaces in a Kruskal diagram can be parametrically determined by fixing the $\tau$ parameter in Eqs.~(\ref{V_(tau, r)_BH}) and (\ref{U_(tau, r)_BH}). 

Furthermore, since Eqs.~(\ref{V_(tau, r)_BH}) and (\ref{U_(tau, r)_BH}) in fact serve as the coordinate transformations from $(\tau, r)$ to $(V, U)$,  the trajectories of the corresponding geodesics are given by substituting Eq.~(\ref{tau_r_relation}) into Eqs.~(\ref{V_(tau, r)_BH}) and (\ref{U_(tau, r)_BH}) to replace $\tau$ as
\begin{equation}\label{V_geodesics}
V(r)=e^{\sqrt{p}(A(r)+\tau_0)/4M}e^{-S(r)/4M}
\end{equation}
and 
\begin{equation}\label{U_geodesics}
U(r)=e^{r/2M}\left(1-\frac{r}{2M}\right)e^{-\sqrt{p}(A(r)+\tau_0)/4M}e^{S(r)/4M},
\end{equation}
where $A(r)$ is defined in Eq.~(\ref{tau_r_relation}). Notice that besides the integral constant $\tau_0$, there is another adjustable integral constant $C$ hidden in $S(r)$ through Eq.~(\ref{general_tau(t,r)_in_S_r*}). We will discuss the choices of those constants later when we connect the infalling geodesics to the outgoing geodesics emitted from the white hole. Also, since the trajectories of geodesics given by Eqs.~(\ref{V_geodesics}) and (\ref{U_geodesics}) are parameterized by $r$ instead of $\tau$,  the corresponding tangent vectors are not the four-velocities of observers moving along those geodesics. To derive the four-velocity, one has to use Eqs.~(\ref{V_(tau, r)_BH}) and (\ref{U_(tau, r)_BH}) to have
\begin{equation}\label{four-velocity_V(tau_r)}
\frac{d V}{d \tau}=\frac{\partial V}{\partial \tau}+\frac{\partial V}{\partial r}\frac{d r}{d \tau}=\frac{V}{4Mf}\left(E-\sqrt{E^2-f}\right), 
\end{equation}
and
\begin{equation}\label{four-velocity_U(tau_r)}
\frac{d U}{d \tau}=\frac{\partial U}{\partial \tau}+\frac{\partial U}{\partial r}\frac{d r}{d \tau}=\frac{-U}{4Mf}\left(E+\sqrt{E^2-f}\right), 
\end{equation}
where $\sqrt{p}=1/E$ is used to write the final expression in terms of $E$. Notice that at the event horizon, where $r=2M$ and $U=0$, the four-velocity $(dV/d\tau, dU/d\tau)$ is well-defined. One can check this by using Eq.~(\ref{Kruskal_UV(r)}) to replace $U$ in the above expression first, and then take the limit $f\rightarrow 0 $ to have\footnote{The derivation of Eq.~(\ref{four_velocity_at_horizon}) and some related relations are given in Appendix~\ref{Sec: four_velocity_in_KS}. }
\begin{equation}\label{four_velocity_at_horizon}
\left( \frac{d V}{d \tau}, \frac{d U}{d \tau}\right)=\left( \frac{V}{8ME}, \frac{eE}{2MV}\right).
\end{equation}
This form shows that the four-velocity is well-defined with only one exception at $V=0$, \textit{i.e.}, when both $U=V=0$.


\subsection{The outgoing radial geodesics from the white hole }\label{SubSec:radial_outgoing_in_Null_KS}

With the formalism related to infalling radial geodesics in the previous two sections, we can obtain the relations of the outgoing radial geodesics emitted from a white hole similarly. For those relations containing the $\epsilon$ factor, the only change is that we choose $\epsilon=-1$ for the outgoing geodesics this time. 
Next, we notice that the exponents in Eq.~(\ref{Null_KS_in_r_t}) for the outgoing radial geodesics become
\begin{equation}\label{t+r^{*}_WH}
t+r^{*}=\sqrt{p}\tau+S(r)+2r^{*}; \quad t-r^{*}=\sqrt{p}\tau+S(r). 
\end{equation}
Also, the regions considered now are regions $I$ and $IV$ in Fig.~\ref{fig:Simple_KS}, so the signs in Eq.~(\ref{Null_KS_in_r_t}) have to be chosen accordingly. Thus we have the transformation relations between $(\tau, r)$ and $(V,U)$ as           
\begin{equation}\label{V_(tau, r)_WH}
V(\tau, r)=-e^{r/2M}\left(1-\frac{r}{2M}\right)e^{\sqrt{p}\tau/4M}e^{S/4M},
\end{equation}
and 
\begin{equation}\label{U_(tau, r)_WH}
U(\tau, r)=-e^{-\sqrt{p}\tau/4M}e^{-S/4M}, 
\end{equation}
which can be compared with their infalling black hole counterparts to see the difference. By substituting Eq.~(\ref{tau_r_relation}) with $\epsilon=-1$ into Eqs.~(\ref{V_(tau, r)_WH}) and (\ref{U_(tau, r)_WH}), we can obtain the trajectories of the outgoing geodesics as
\begin{equation}\label{V_geodesics_WH}
V(r)=-e^{r/2M}\left(1-\frac{r}{2M}\right)e^{-\sqrt{p}(A(r)-\tau_0)/4M}e^{S(r)/4M},
\end{equation}
and 
\begin{equation}\label{U_geodesics_WH}
U(r)=-e^{\sqrt{p}(A(r)-\tau_0)/4M}e^{-S(r)/4M}.  
\end{equation}


\subsection{Matching the outgoing and infalling radial geodesics}\label{SubSec:matching_in/out_Null_KS}

\begin{figure}[h!]
	\centering
	\includegraphics[scale=0.5]{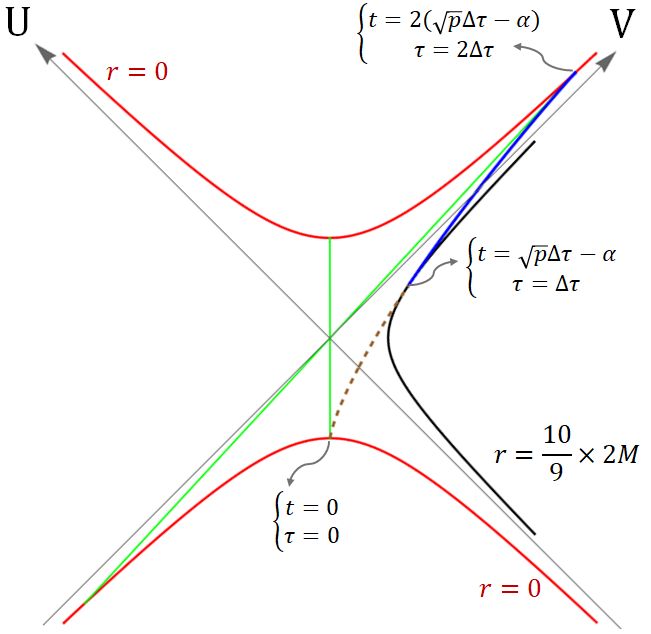}
	\caption{Matching the outgoing part (the brown dashed line) to the infalling part (the blue solid line) of a bounded radial geodesic at the maximal radius it can reach. The constants of integration related to the ingoing and outgoing part, $\tau_{out/in}$ and $C_{out/in}$, are fully determined once the initial position is specified, \textit{i.e.} the values of $t$ and $\tau$ at $r=0$ of the white hole part. }  
	\label{fig:matching_in_KS}
\end{figure}

In the maximally extended Schwarzschild spacetime, a bounded outgoing geodesic should connect to a specific infalling geodesic at the given maximal radius $R$ to form a single bounded radial geodesic. The way to match the outgoing and infalling parts can be determined by adjusting the constants of integration $\tau_0$ and $C$ of them as follows. From here, we distinguish those constants of outgoing and infalling parts of a radial geodesic by labeling them as $\tau_{out/in}$ and $C_{out/in}$ accordingly. We first notice that the function $A(r)$ from Eq.~(\ref{tau_r_relation}) at $r=R$ and $r=0$ are given by 
\begin{equation}\label{A(r)_BC}
\begin{split}
    A(R)&=0,  \\ 
    A(0)&=\frac{R}{2}\sqrt{\frac{R}{2M}}\pi . 
\end{split}
\end{equation}
Thus the total elapsed proper time $\Delta \tau$ related to the infalling or outgoing part of a bounded radial geodesic is given by $\Delta \tau=A(R)-A(0)=A(R)$. Then, to set the initial condition that $\tau=0$ at $r=0$ of the white hole part, one has to set $\tau_{out}=\Delta\tau$. 

Next, by using Eq.~(\ref{general_tau(t,r)}), the other constant of integration $C_{out}$ is then related to the starting ``position" of an outgoing geodesic, since the coordinates $t$ and $r$ switch their characters inside a black or white hole. Using the function $\Theta(r)$ at $r=R$ and $r=0$,
\begin{equation}\label{Theta_BC}
\begin{split}
    \Theta(R)&=0, \\
    \Theta(0)&=\frac{\pi M\left(\frac{4M}{R}-1\right)}{\sqrt{\left(1-\frac{2M}{R}\right)\frac{2M}{R}}} \equiv \alpha,      
\end{split}
\end{equation}
we can see that to select the radial geodesic starting from  $t=t_{i}$ at the initial singularity ($r=0$ of the white hole part), we have to choose $C_{out}=t_{i}-\alpha$. 

Once we have determined the two constants for the outgoing part of a bounded radial geodesic, we can use those two values and Eq.~(\ref{general_tau(t,r)}) to determine ``when" this geodesic reaches maximal radius $R$ measured in terms of $t$, which serves the initial condition for the infalling part. Then one can use the same method to determine the correct integral constants $\tau_{in}$ and $C_{in}$ for the infalling part. The trajectory of a bounded radial geodesic determined through this method is shown in Fig. \ref{fig:matching_in_KS}.   


\section{The Bounded Radial Geodesics in the Black-to-white Hole Transition}\label{Sec:Bound_geodesics_matching}

\begin{figure}[h!]
	\centering
	\includegraphics[scale=0.5]{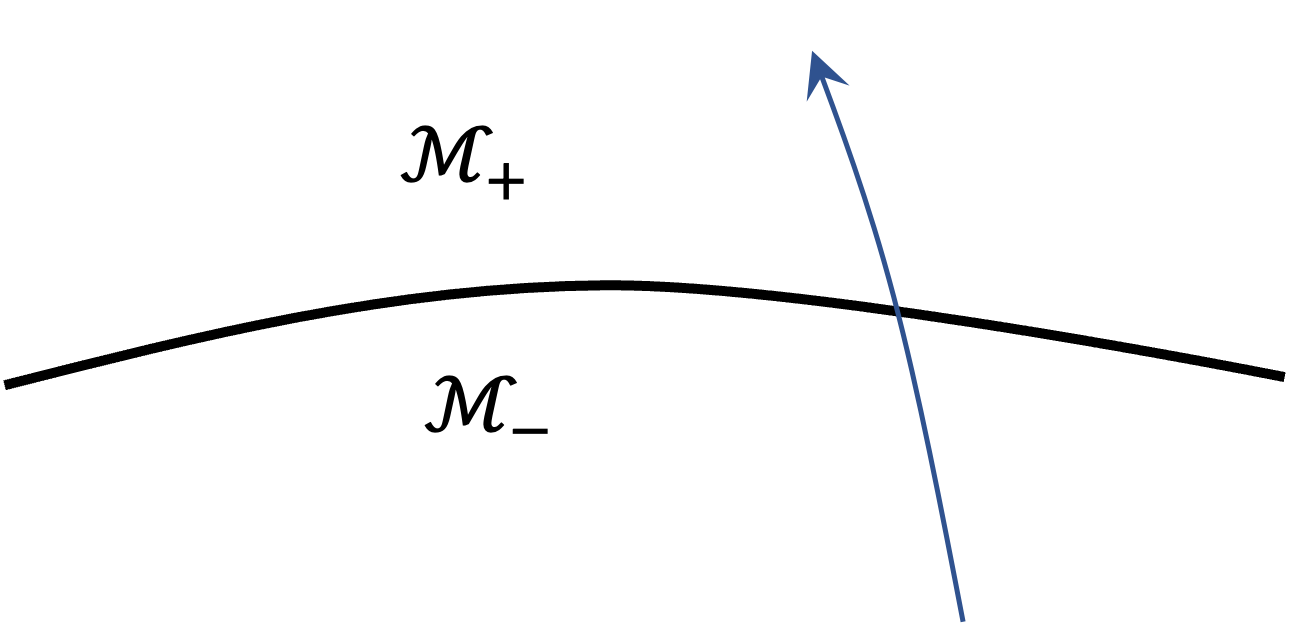}
	\caption{The convention of the subscript $\pm$ used in this work is that the timelike geodesics cross the shell from  $\mathcal{M}_{-}$ to $\mathcal{M}_{+}$. }  
	\label{fig:+&-_convention}
\end{figure}

Now we extend the discussion of the bounded radial geodesics of the most extended Schwarzschild spacetime to those of the generalized black-to-white hole bouncing models, in which two Schwarzschild solutions with different mass parameters $M_{\pm}$ are connected by an effective static thin shell inside the event horizon \cite{Hong:2022thd}. That is, on each side of the shell, the metric ansatz is given by 
\begin{equation}\label{Schwarzschild_metric_inside_BWH}
d s_{\pm}^2=-(-f_{\pm})^{-1}d r_{\pm}^2+(-f_{\pm})d t_{\pm}^2+r_{\pm}^2d\Omega^2, 
\end{equation}
which describes the metric inside the horizon, with
\begin{equation}\label{f(r)_function}
f_{\pm} = 1-\frac{2M_{\pm}}{r_{\pm}},
\end{equation}
where the labels $-$ and $+$ are used to distinguish the quantities of the spacetimes (phases) separated by the shell with the convention that radial geodesics move from  $\mathcal{M}_{-}$ to $\mathcal{M}_{+}$ as shown in Fig. \ref{fig:+&-_convention}. 
Meanwhile, Eq.~(\ref{Schwarzschild_metric_inside_BWH}) is nothing but the usual Schwarzschild metric (\ref{Schwarzschild_metric}) expressed in the form emphasizing the changing the roles of the coordinates $\{t, r\}$ inside the event horizon. Furthermore, for the special case in which the shell is static, \textit{i.e.} $r_{+}=r_{-}=b$, it was shown that the first junction condition gives the following relation
\begin{equation}\label{1JC_BWH_t_component}
\sqrt{-f_{-}(b)}d t_{-}=\sqrt{-f_{+}(b)}d t_{+},  
\end{equation}
which further leads to the energy shift relation of a radial geodesic crossing the shell \cite{Hong:2022thd}
\begin{equation}\label{energy_shift_BWH}
E_{+}=\sqrt{\frac{f_{+}(b)}{f_{-}(b)}}E_{-},   
\end{equation}
and also the relation between the $t_{\pm}$ coordinates in the two phases separated by the shell \cite{Lin:2023ztq}
\begin{equation}\label{t_relation_BWH}
t_{+}=\sqrt{\frac{f_{-}(b)}{f_{+}(b)}}t_{-}. 
\end{equation}

Meanwhile, the Kruskal diagram is generally unsuitable for us to visualize the spacetime constructed via the thin-shell approximation.  Typically, the Penrose diagram is used in such a scenario. However, unlike the usual vanilla transformation converting the null KS coordinates to the compactified coordinates of the Penrose diagram $(\tilde{V}, \tilde{U})$:
\begin{equation}\label{KS_to_Penrose}
(\tilde{V}, \tilde{U})=(\tan^{-1}V, \tan^{-1}U), 
\end{equation}
a ``good'' Penrose diagram of the above-mentioned model cannot be constructed simply by this transformation. The general issue of the Penrose diagrams of the spherically symmetric spacetime models constructed via a static spacelike thin shell is studied in Ref.~\cite{Lin:2023ztq}. In a Penrose diagram constructed through a simple cut-and-paste procedure, an implicit discontinuity exists at the thin shell due to the geometric property of Penrose diagrams in general, so an extra transformation is required except in some special cases.  However, this extra transformation unavoidably re-introduces a special type of coordinate singularity back to the event horizon. Here we quickly review the procedure given in Ref.~\cite{Lin:2023ztq}, and show how one  identifies radial geodesics in the two phases separated by the shell after this procedure. After establishing the trajectories of the radial geodesics in the resulting Penrose diagram, we then show that they are indeed well-behaved in the resulting Penrose diagram, but they degenerate to the trajectory of null geodesics at the event horizon due to the above-mentioned singularity.


\subsection{The cut-and-paste procedure generating a Penrose diagram without illness at the thin shell}\label{subsec:cut_and_paste_procedure_BWH}

\begin{figure}[h!]
	\centering
	\includegraphics[scale=0.2]{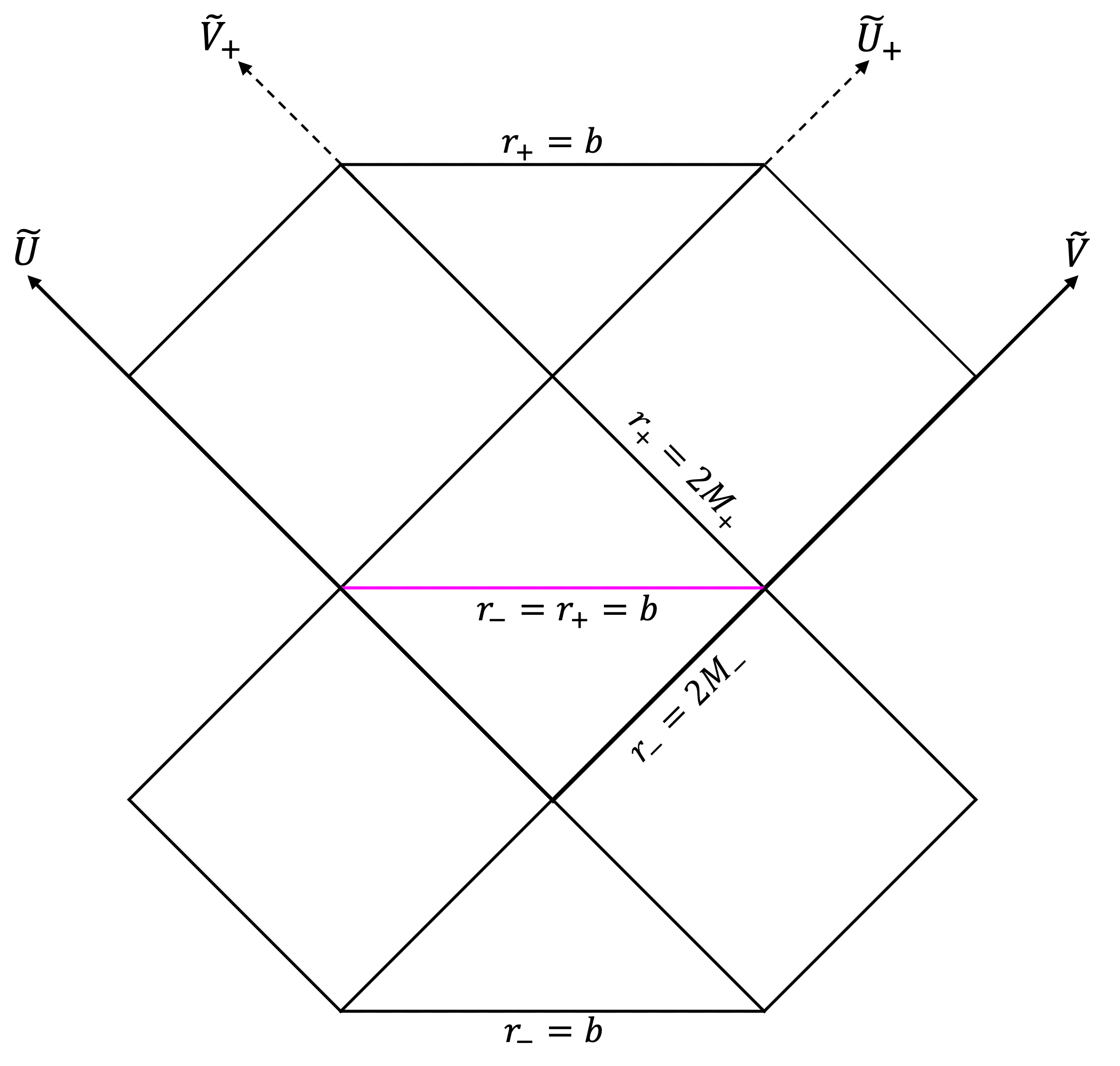}
	\caption{The connected Penrose diagram for the generalized black-to-white hole bounce.}  
	\label{fig:BWH_trans}
\end{figure}

 Starting from the null KS-coordinates $(V_{\pm}, U_{\pm})$ for the Schwarzschild solution on each side of the shell, one first applies the following rescaling transformation   
\begin{equation}\label{ModKS}
(V'_{\pm}, U'_{\pm})\equiv \left(\frac{V_{\pm} }{X_{\pm}},  \frac{U_{\pm} }{X_{\pm}}\right), 
\end{equation}
where $X_{\pm} \equiv \sqrt{1-b/2M_{\pm}}e^{b/4M_{\pm}}$. In the new coordinates, the trajectory of the shell is then given by $V'_{\pm} U'_{\pm}=1$, which will become a straight line after compactification by using Eq.~(\ref{KS_to_Penrose}).  Also, notice that before and after the rescaling transformation, we have
\begin{equation}\label{U/V_mini}
\frac{U'_{\pm}}{V'_{\pm}}=\frac{U_{\pm}}{V_{\pm}}= e^{-t_{\pm}/2M_{\pm}},
\end{equation}
which implicitly indicates that the result of matching the radial geodesics on two sides of the shell is incompatible with the relation given by the first junction condition, Eq.~(\ref{t_relation_BWH}). To fix the discontinuity at the thin shell, \textit{i.e.} to enforce Eq.~(\ref{t_relation_BWH}) in the resulting Penrose diagram, 
one performs a second conformal transformation on $(V'_{+}, U'_{+})$, 
\begin{equation}\label{Second_conformal_BHtoWH}
\left(V''_{+}, U''_{+}\right)=\left(V'_{+} \left| V'_{+} \right| ^{k-1}, U'_{+} \left| U'_{+} \right| ^{k-1} \right),
\end{equation}
where $k \equiv \frac{M_{+}}{M_{-}}\sqrt{f_{+}(b)/f_{-}(b)}$
such that 
\begin{equation}\label{second_conformal_moving_t}
\frac{U''_{+}}{V''_{+}}=\pm e^{-\frac{t_{+}}{2M_{-}}\sqrt{f_{+}(b)/f_{-}(b)}}. 
\end{equation}
After the above two transformations, one then can compactify $(V'_{-}, U'_{-})$ and $(V''_{+}, U''_{+})$ by using the inverse tangent function such that
\begin{equation}
\begin{split}
(\tilde{V}_{-}, \tilde{U}_{-})=&\left(\tan^{-1}V'_{-}, \tan^{-1}U'_{-} \right),  \\
(\tilde{V}_{+}, \tilde{U}_{+})=&\left(\tan^{-1}V''_{+}, \tan^{-1}U''_{+} \right),
\end{split}
\end{equation}
and then cut out the unwanted part $r_{\pm}<b$. Lastly, to paste the remaining part together, one uses the following identifications 
\begin{equation}\label{identifaction_1st_BH}
(\tilde{V}, \tilde{U})  \equiv (\tilde{V}_{-}, \tilde{U}_{-}),  
\end{equation}
and
\begin{equation}\label{identification_2nd_BH}
(\tilde{V}, \tilde{U}) =
     \left(\tilde{U}_{+}+\frac{\pi}{2}, \tilde{V}_{+}+\frac{\pi}{2}\right).    
\end{equation}
Then, we obtain the resulting Penrose diagram without illness at the thin shell, Fig~\ref{fig:BWH_trans}. Meanwhile, the coordinates of the resulting Penrose diagram $(\tilde{V}, \tilde{U})$ also form a coordinate chart covering the entire thin shell.


\subsection{The bounded radial geodesics}\label{subsec: bounded_radial_geodesic_BWH}

\begin{figure}[h!]
	\centering
	\includegraphics[scale=0.3]{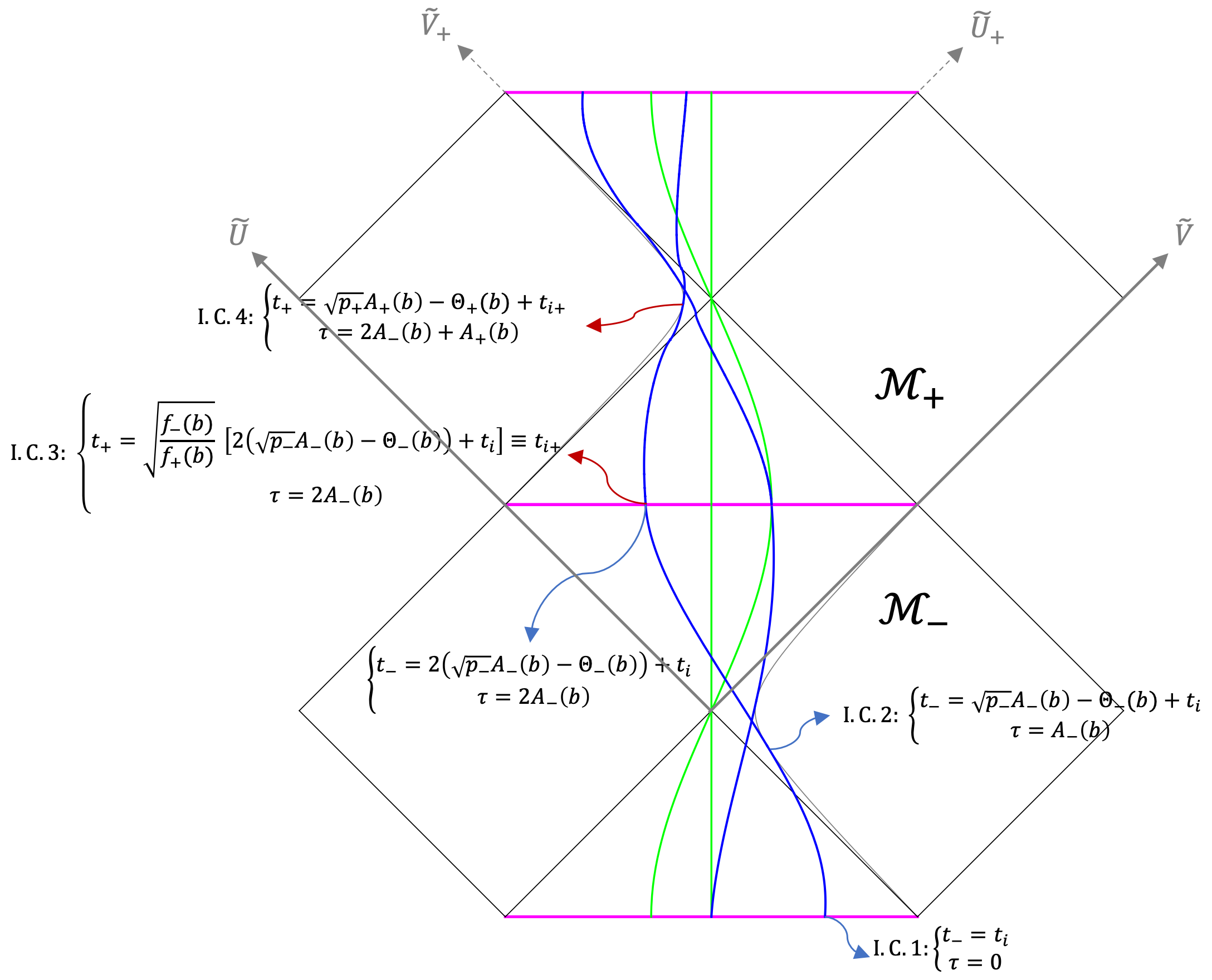}
	\caption{The trajectories of the bounded radial geodesics in the Penrose diagram of the generalized black-to-white hole bounce with $M_{+}>M_{-}$. The blue lines are trajectories of bounded radial geodesics with some initial energy parameter $E_{-}$, while the green lines are the special radial geodesics with $E_{-}=0$. The initial conditions of each segment of one bounded radial geodesic are shown.  Notice that after the initial condition for the first segment is determined (labeled as $I.C.1$ in the diagram), the initial conditions for the rest of the segments ($I.C.2-4$) have to be chosen accordingly. \\ Next, the smoothness of those trajectories at the thin shell, Eq.~(\ref{Smoothly_crossing_condition}), is consistent with the energy shift condition Eq.~(\ref{energy_shift_BWH}). Then reversely, this resulting Penrose diagram is healthy at the thin shell, \textit{i.e.} without implicit discontinuity nor artificial cusp for the radial geodesics shown in this diagram. Thus, the coordinates $(\tilde{V}, \tilde{U})$ serve as a well-behaved coordinate chart covering the entire thin shell and its neighborhood. However, the second transformation Eq.~(\ref{Second_conformal_BHtoWH}) introduces a new type of coordinate singularity back to the event horizons in $M_{+}$, which is manifested by the distortion of the radial geodesics (blue) at the very place. Since $M_{+}>M_{-}$, this distortion is of the stretching type (the radial geodesics become parallel to the event horizon when crossing it) \cite{Lin:2023ztq}.}  
	\label{fig:BWH_geodesics}
\end{figure}

Since the metrics of the two phases separated by the thin shell are of the Schwarzschild form, the trajectory of a bounded radial geodesic, $\left(V_{\pm}(r_{\pm}), U_{\pm}(r_{\pm})\right)$, is still given by Eqs.~(\ref{V_geodesics}) and (\ref{U_geodesics}) for the infalling part, and Eqs.~(\ref{V_geodesics_WH}) and (\ref{U_geodesics_WH}) for the outgoing part. Nevertheless, we have to choose the correct constants of integration $\tau_{(out/in)(\pm)}$ and $C_{(out/in)(\pm)}$ segment by segment for any bounded radial geodesic as shown in Fig.~\ref{fig:BWH_geodesics}.

For the first phase labeled as $\mathcal{M}_{-}$, the procedure to determine the four constants of integration $\tau_{(out/in)-}$ and $C_{(out/in)-}$ is similar to what we have done in Sec. \ref{SubSec:matching_in/out_Null_KS}. Here we choose that all of the geodesics start from the bottom of the diagram, \textit{i.e.} $r_{-}=b$ of the white hole part. Then, for a bounded radial geodesic starting from the ``position" $t_{-}(r_{-}=b)=t_{i}$ with $\tau(r_{-}=b)=0$, by using Eqs.~(\ref{tau_r_relation}) and (\ref{general_tau(t,r)}), one has to choose the corresponding constants as 
\begin{equation}\label{1st_WH_constants}
\begin{split}
\tau_{out-}&=A_{-}(b), \\
C_{out-}&=t_{i}-\Theta_{-}(b), 
\end{split}
\end{equation}
where the subscript ``$-$'' for both the functions $A_{-}(r)$ and $\Theta_{-}(r)$ is required, since their values are also related to the energy parameter corresponding to the bounded radial geodesic as $E^2_{-}=1-2M_{-}/R_{-}=1/p_{-}$. With the constants $\tau_{out-}$ and $C_{out-}$, we can use Eqs.~(\ref{tau_r_relation}) and (\ref{general_tau(t,r)}) again to show that this bounded radial geodesic reaches the maximal radius $r_{-}=R_{-}$ at 
\begin{equation}\label{first_ConnectPoint_BWH}
\begin{split}
\tau&=A_{-}(b) \\
t_{-}&=\sqrt{p_{-}}A_{-}(b)-\Theta_{-}(b)+t_{i}. 
\end{split}
\end{equation}
Using those two conditions, we can obtain the two constants related to the infalling segment in the region $\mathcal{M}_{-}$ as
\begin{equation}\label{1st_BH_constants}
\begin{split}
\tau_{in-}&=A_{-}(b), \\
C_{in-}&=-t_{i}+\Theta_{-}(b).
\end{split}
\end{equation}
With the values of the two constants $\tau_{in-}$ and $C_{in-}$, by repeating the similar procedure, we can determine where the geodesic crosses the thin shell in terms of the coordinates $(t_{-}, r_{-})$ as 
\begin{equation}
t_{-}(r_{-}=b)=2\left(\sqrt{p_{-}}A_{-}(b)-\Theta_{-}(b)\right)+t_{i}, 
\end{equation}
with the elapsed proper time $\tau=2A(b)$. This point, or say the spacetime event, is also the point where we connect the first infalling segment to the second outgoing segment. Naturally, the elapsed proper time $\tau=2A(b)$ for the starting point of the second outgoing segment is unchanged. However, the starting position for the second outgoing segment must be given in terms of the coordinates $(t_{+}, r_{+})$. From Eq.~(\ref{t_relation_BWH}), we can show that it is given by 
\begin{equation}\label{BWH_t_connecting}
t_{+}(r_{+}=b)=\sqrt{\frac{f_{-}(b)}{f_{+}(b)}}\left[2\left(\sqrt{p_{-}}A_{-}(b)-\Theta_{-}(b)\right)+t_{i}\right]. 
\end{equation}
Using Eq.~(\ref{BWH_t_connecting}) and $\tau=2A(b)$ as the initial condition for the second outgoing segment, we then can determine the two constants for this segment as
\begin{equation}\label{2st_WH_constants}
\begin{split}
\tau_{out+}&=2A_{-}(b)+A_{+}(b), \\
C_{out+}&=\sqrt{\frac{f_{-}(b)}{f_{+}(b)}}\Big(t_{i}-2\Theta_{-}(b)\Big)-\Theta_{+}(b).
\end{split}
\end{equation}
Repeating the repetitive procedure used for the first phase $\mathcal{M}_{-}$, the constants for the second infalling part are given by
\begin{equation}\label{2st_BH_constants}
\begin{split}
\tau_{in+}&=2A_{-}(b)+A_{+}(b), \\
C_{in+}&=-\sqrt{\frac{f_{-}(b)}{f_{+}(b)}}\Big(t_{i}-2\Theta_{-}(b)\Big)+\Theta_{+}(b).
\end{split}
\end{equation}
An example of the resulting Penrose diagram with trajectories of bounded radial geodesics is given in Fig. \ref{fig:BWH_geodesics}.


\subsection{Deriving the energy shift relation in a coordinate dependent way}\label{Subsec:derive_energy_shift_BWH}

Now we show that the energy shift relation, Eq.~(\ref{energy_shift_BWH}), is consistent with the condition that a timelike radial geodesic crosses the thin shell smoothly in the Penrose diagram, \textit{e.g.} Fig. \ref{fig:BWH_geodesics}. That is, the condition
\begin{equation}\label{Smoothly_crossing_condition}
\frac{d\tilde{U}}{d\tilde{V}} \Big|_{r_{-} \to b}=\frac{d\tilde{U}}{d\tilde{V}} \Big|_{r_{+} \to b}
\end{equation}
gives Eq.~(\ref{energy_shift_BWH}).

We first notice that for the black hole part, the radial infalling geodesics given by Eqs.~(\ref{V_geodesics}) and (\ref{U_geodesics}) can be re-expressed into the following form
\begin{equation}\label{V_U_minus_exp_form}
(V_{-}, U_{-})=\left(e^{g_{-}(r_{-})}, e^{h_{-}(r_{-})}\right), 
\end{equation}
where 
\begin{equation}\label{g_{-}(r)}
g_{-}(r_{-})=\frac{\sqrt{p}}{4M}\left(A(r)-\tau_0\right)-\frac{\Theta(r)-r^{*}+C}{4M}, 
\end{equation}
and 
\begin{equation}\label{h_{-}(r)}
h_{-}(r_{-})=\frac{r}{2M}+\log\left(1-\frac{r}{2M}\right)-g_{-}(r),
\end{equation}
respectively. And notice that all symbols on the r.h.s. should carry the subscript ``$-$'', but we neglect it for the cleanness.  Using Eq.~(\ref{V_U_minus_exp_form}), the trajectory equation of a radial infalling geodesic is given by  
\begin{equation}
(\tilde{V}, \tilde{U})= \left(\tan^{-1}\frac{e^{g_{-}}}{X_{-}}, \tan^{-1}\frac{e^{h_{-}}}{X_{-}}\right).
\end{equation}
By using the following differential relations
\begin{equation}\label{differential relations_A_Theta_r*}
\begin{split}
\frac{dA(r)}{dr}&=\frac{-1}{\sqrt{E^2-f}}=\frac{-\sqrt{p}}{\sqrt{1-pf}},   \\
\frac{d\Theta(r)}{dr}&= \frac{\sqrt{1-pf}}{f}, \\
\frac{dr^{*}}{dr}&=\frac{r}{r-2M},
\end{split}
\end{equation}
we then have
\begin{equation}\label{dg-/dr}
\begin{split}
\frac{dg_{-}}{dr}=& \frac{1}{4M}\left(\sqrt{p}\frac{dA(r)}{dr}-\frac{d\Theta(r)}{dr}+\frac{dr^{*}}{dr} \right) \\
=& \frac{1}{4M}\left(\frac{-p}{\sqrt{1-pf}}-\frac{\sqrt{1-pf}}{f}+\frac{r}{r-2M} \right) \\
=& \frac{1}{4M}\left(\frac{\sqrt{1-pf}-1}{f\sqrt{1-pf}} \right),
\end{split}
\end{equation}
and 
\begin{equation}\label{df-/dr}
\begin{split}
\frac{dh_{-}}{dr}&=\frac{1}{2Mf}-\frac{1}{4M}\left(\frac{\sqrt{1-pf}-1}{f\sqrt{1-pf}} \right)\\
&=\frac{1}{4M}\left(\frac{\sqrt{1-pf}+1}{f\sqrt{1-pf}} \right).
\end{split}
\end{equation}
Thus, from the black hole side we have
\begin{equation}\label{dU/dV_BH}
\frac{d\tilde{U}}{d\tilde{V}}=\frac{d\tilde{U}/dr}{d\tilde{V}/dr}=\frac{1+e^{2g_{-}}/X_{-}^2}{1+e^{2h_{-}}/X_{-}^2} \times e^{(h_{-}-g_{-})} \times \frac{dh_{-}/dr}{dg_{-}/dr}=\frac{1+\tan^{2}\tilde{V}}{1+\tan^{2}\tilde{U}} \times \frac{U_{-}}{V_{-}} \times \frac{\sqrt{1-p_{-}f_{-}}+1}{\sqrt{1-p_{-}f_{-}}-1}. 
\end{equation}

Next, crossing the thin shell to the white hole part, the radial geodesics become outgoing and their trajectories are given by Eqs.~(\ref{V_geodesics_WH}) and (\ref{U_geodesics_WH}). By a change of the constant $\tau_0=-\tau'_0$, these two equations have the form of exchanging their black hole counterpart, Eqs.~(\ref{V_geodesics}) and (\ref{U_geodesics}), with an extra minus sign. Thus, for the radially outgoing geodesics we have 
\begin{equation}\label{V_U_plus_exp_form}
(V_{+}, U_{+})=\left(-e^{h_{+}(r_{+})}, -e^{g_{+}(r_{+})}\right),
\end{equation}
where $g_{+}(r_{+})$ and $h_{+}(r_{+})$ are given by Eqs.~(\ref{g_{-}(r)}) and (\ref{h_{-}(r)}) respectively, with the subscript ``$-$'' replaced by ``$+$''. Then, using Eq.~(\ref{V_U_plus_exp_form}), the trajectory in the connected Penrose diagram is given by 
\begin{equation}
(\tilde{V}, \tilde{U})=\left(\tilde{U}_{+}+\frac{\pi}{2}, \tilde{V}_{+}+\frac{\pi}{2}\right)= \left(\tan^{-1}[X^k_{+}e^{-kg_{+}}], \tan^{-1}[X^k_{+}e^{-kh_{+}}]\right), 
\end{equation}
where $k = \frac{M_{+}}{M_{-}}\sqrt{f_{+}(b)/f_{-}(b)}$ is from the second conformal transformation Eq.~(\ref{Second_conformal_BHtoWH}).  
With the similar calculation where the details are given in Appendix \ref{Sec: the derivation}, the slope of a radially outgoing geodesic in the $\mathcal{M}_{+}$ region of Fig. \ref{fig:BWH_geodesics} is given by 
\begin{equation}\label{dU/dV_WH}
\frac{d\tilde{U}}{d\tilde{V}}=\frac{1+\tan^{2}\tilde{V}}{1+\tan^{2}\tilde{U}} \times \frac{U''_{+}}{V''_{+}} \times \frac{\sqrt{1-p_{+}f_{+}}+1}{\sqrt{1-p_{+}f_{+}}-1}. 
\end{equation}
Then, by substituting Eqs.~(\ref{dU/dV_BH}) and (\ref{dU/dV_WH}) into Eq.~(\ref{Smoothly_crossing_condition}), the condition that the trajectory of a radial geodesic shown in the connected Penrose diagram crosses the thin shell smoothly, we arrive the result 
\begin{equation}\label{smoothness_result}
p_{-}f_{-}(b)=p_{+}f_{+}(b),
\end{equation}
where Eqs.~(\ref{U/V_mini}), (\ref{second_conformal_moving_t}) and (\ref{t_relation_BWH}) are used. By noticing that $p_{\pm}=1/E^2_{\pm}$, we see that Eq.~(\ref{smoothness_result}) is nothing but the energy shift relation, Eq.~(\ref{energy_shift_BWH}). Notice that this energy shift relation is derived in a coordinate-independent way in Ref. \cite{Hong:2022thd}. This means that the coordinates of the connected Penrose diagram $(\tilde{V}, \tilde{U})$ form a well-behaved coordinate chart covering the entire thin shell in the sense that the trajectories of radial geodesics do not have artificial cusp nor implicit discontinuity around the thin shell in this coordinate system.       


\subsection{The illness of the coordinates at the horizon }\label{Subsec:the_illness_at_horizon}

\begin{figure}[h!]
	\centering
	\includegraphics[scale=0.7]{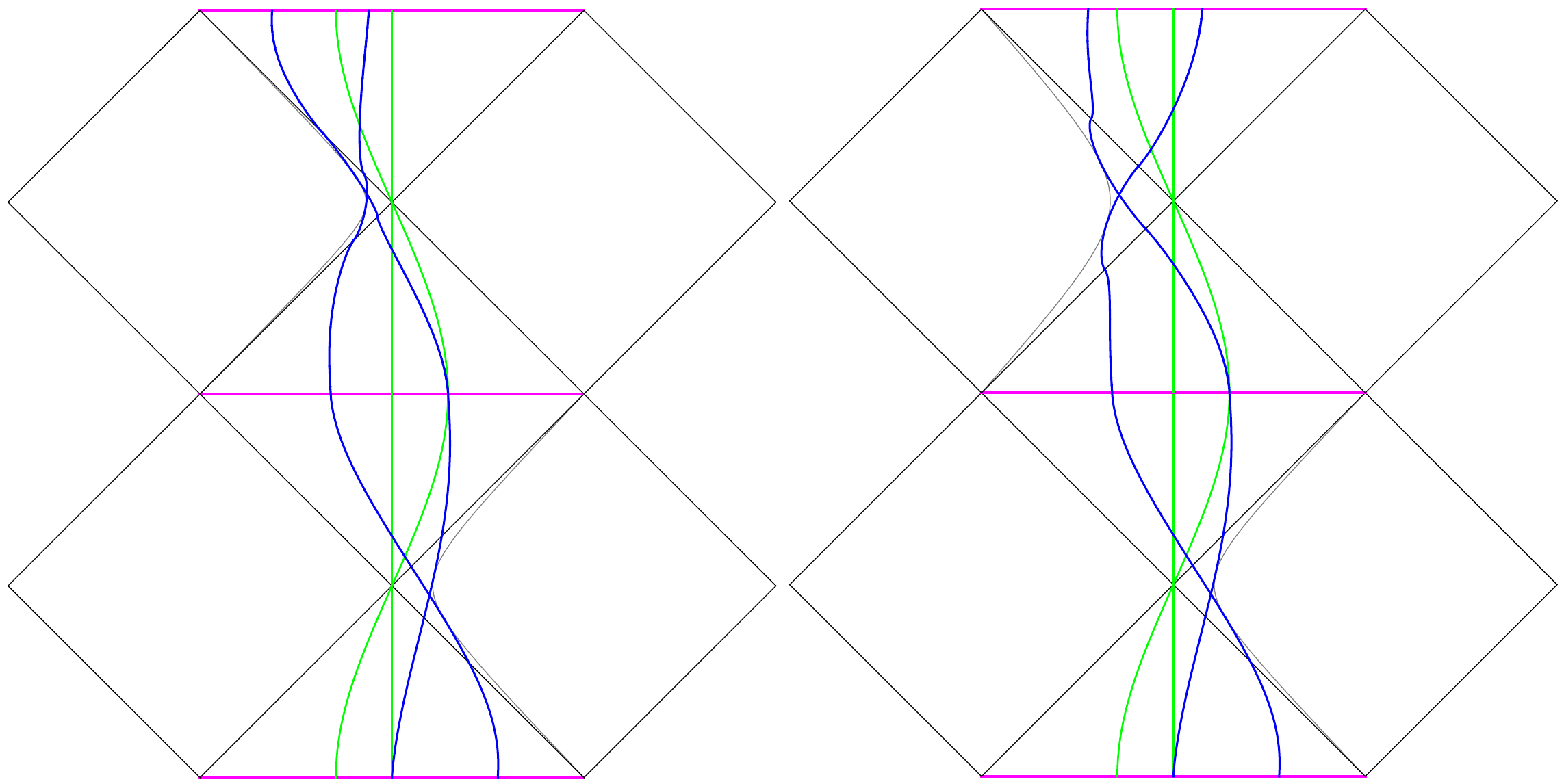}
	\caption{
 The diagram on the left is a mass-increasing scenario with $M_{+}/M_{-}=1.2$, while the diagram on the right is a mass-decreasing scenario with $M_{+}/M_{-}=0.8$. Notice that the bottom halves of the two diagrams are the same, but the second halves on the top of them are different. The distortion can be distinguished into the stretching type (the left figure) and the squeezing type (the right figure). See Ref.~\cite{Lin:2023ztq} for the geometric explanation.}   
	\label{fig:second_trans_BHtoWH}
\end{figure}

However, as pointed out in Ref.~\cite{Lin:2023ztq}, the second transformation fixing the implicit discontinuity at the thin shell unavoidably brings back a new type of coordinate singularity to the event horizons. After the second transformation, in terms of the modified KS coordinates Eq.~(\ref{Second_conformal_BHtoWH}), the metric is given by  
\begin{equation}\label{metric_after_2nd_trans}
d s_{+}^2=-\frac{16M_{+}^3}{r_{+}}e^{-r_{+}/2M_{+}}\frac{X_{+}^2}{k^2} \left|V''_{+}U''_{+}\right|^{\frac{1-k}{k}} (d V''_{+} d U''_{+}+d U''_{+} d V''_{+})+r_{+}^2 d \Omega^2, 
\end{equation}
in which the factor $\left|V''_{+}U''_{+}\right|^{\frac{1-k}{k}}$ makes the metric singular at the event horizons, $U''=0$ or $V''=0$. When $k < 1$, \textit{i.e.} $M_{+} < M_{-}$, the metric components $g_{U''_{+}V''_{+}}$ and $g_{V''_{+}U''_{+}}$ are zero at the event horizons. On the other hand, when $k>1$, \textit{i.e.} $M_{+} > M_{-}$, the metric components $g_{U''_{+}V''_{+}}$ and $g_{V''_{+}U''_{+}}$ are divergent at the event horizons. 

The two different forms of this new type of singularity at the event horizons are manifested in the distinct behaviors of the trajectories of radial geodesics in the connected Penrose diagram. In the connected Penrose diagram with $k < 1$, the radial geodesics become perpendicular to the event horizons when crossing them, while in the diagram with $k > 1$, the radial geodesics become parallel to the event horizons at the very places as shown in Fig.~\ref{fig:second_trans_BHtoWH}. In the following, we will demonstrate those two results by showing that the quantity $d\tilde{U}/d\tilde{V}$ of any timelike radial geodesic either approach to zero or become divergent at the event horizons in the upper half of the connected Penrose diagram, \textit{i.e.} the part experienced the second transformation Eq.~(\ref{Second_conformal_BHtoWH}). We will also refer to this quantity as the ``slope'' of a trajectory due to the fact that this name is well-justified after rotating the Penrose diagram $45\degree$ clockwisely. Notice that in the usual Penrose diagram, the slope of any timelike trajectory approaches zero or infinite only when the tangent vector of it approaches the speed of light. We will show that this rule, which is also shared by the usual Kruskal diagram, breaks down at the above-mentioned event horizons.

To obtain the slope of a radial geodesic in the connected Penrose diagram, $d\tilde{U}/d\tilde{V}$, we first have to calculate the slope of a radial geodesic in a Kruskal diagram, \textit{i.e.} $dU/dV$. By using Eqs.~(\ref{four-velocity_V(tau_r)_App}) and (\ref{four-velocity_U(tau_r)_App}), we can easily obtain the slope of the infalling part as
\begin{equation}\label{slope_kruskal}
\frac{dU}{dV}=\frac{dU/d\tau}{dV/d\tau}=\frac{-U}{V}\frac{1+\sqrt{1-f/E^2}}{1-\sqrt{1-f/E^2}}=\frac{U}{V}\frac{\sqrt{1-pf}+1}{\sqrt{1-pf}-1}, 
\end{equation}
where $p =1/E^2$ is used to obtain the last equality. In this form, we can see that it is exactly equal to the last two factors in Eq.~(\ref{dU/dV_BH}). From Eq.~(\ref{four_velocity_at_horizon_App}), we have the slope of an infalling radial geodesic with energy per unit mass $E$ at the black hole event horizon $(U=0)$ given by 
\begin{equation}\label{slope_kruskal_Horizon}
\left. \frac{dU}{dV} \right\vert_{r \rightarrow 2M}=\frac{4eE^2}{V^2},    
\end{equation}
which is positive and finite except at the bifurcation surface, \textit{i.e.} $V=U=0$.    
Similarly, by using Eqs~(\ref{Outgoing_four-velocity_V(tau_r)_App}) and (\ref{Outgoing_four-velocity_U(tau_r)_App}), one can derive the slope of an outgoing radial geodesic to be
\begin{equation}\label{slope_kruskal_out}
\frac{dU}{dV}=\frac{U}{V}\frac{\sqrt{1-pf}-1}{\sqrt{1-pf}+1}, 
\end{equation}
which at the white hole event horizon $(V=0)$ reduces to 
\begin{equation}\label{slope_kruskal_Horizon_WH} 
\left. \frac{dU}{dV} \right\vert_{r \rightarrow 2M}=\frac{U^2}{4eE^2}.    
\end{equation}

Now, on the bottom half of the connected Penrose diagram, $\mathcal{M}_{-}$ part in Fig. \ref{fig:BWH_geodesics}, where only the rescaling transformation $(V'_{-}, U'_{-})=(V'_{-}/X_{-}, U'_{-}/X_{-})$ is performed, the slope of the trajectory of a radial geodesic with energy parameter $E_{-}$ at the black hole event horizon $(U_{-}=0)$ is given by Eq.~(\ref{dU/dV_BH}) as
\begin{equation}
\left. \frac{d\tilde{U}}{d\tilde{V}} \right\vert_{r \rightarrow 2M_{-}}=
\left.
\frac{1+\tan^{2}\tilde{V}}{1+\tan^{2}\tilde{U}} \left( \frac{U_{-}}{V_{-}}  \frac{\sqrt{1-p_{-}f_{-}}+1}{\sqrt{1-p_{-}f_{-}}-1}\right)
\right\vert_{r \rightarrow 2M_{-}}=\left(1+\tan^{2}\tilde{V}\right) \frac{4eE_{-}^2}{V_{-}^2},    
\end{equation}
where Eqs.~(\ref{slope_kruskal}) and (\ref{slope_kruskal_Horizon}) are used to obtain the last equality. Similar to the situation in the Kruskal diagram, with $0<E<\infty$, the slope of the trajectory at the horizon is also positive and finite except at the bifurcation surface.   

However, for the upper half in Fig. \ref{fig:BWH_geodesics} labeled by $\mathcal{M}_{+}$, in which an additional coordinate transformation Eq.~(\ref{Second_conformal_BHtoWH}) is performed, the slope of the trajectory of an outgoing radial geodesic at the white hole event horizon $(V_{+}=0)$ is given by Eq.~(\ref{dU/dV_WH}) as\footnote{Since we are utilizing Eq.~(\ref{dU/dV_WH}), this result is given with the condition that both $V_{+}$ and $U_{+}$ are negative.}  
\begin{equation}\label{problem_at_horizon}
\begin{split}
\left. \frac{d\tilde{U}}{d\tilde{V}} \right\vert_{r \rightarrow 2M_{+}}
&=\left.
\frac{1+\tan^{2}\tilde{V}}{1+\tan^{2}\tilde{U}}  \frac{U''_{+}}{V''_{+}}  \frac{\sqrt{1-p_{+}f_{+}}+1}{\sqrt{1-p_{+}f_{+}}-1}
\right\vert_{r \rightarrow 2M_{+}} \\
&=\left.
\frac{1+(-U_{+}/X_{+})^{2k}}{1+(-V_{+}/X_{+})^{2k}}  \left(\frac{V_{+}}{U_{+}} \right)^k \frac{\sqrt{1-p_{+}f_{+}}+1}{\sqrt{1-p_{+}f_{+}}-1}
\right\vert_{r \rightarrow 2M_{+}} \\
&=\left(1+\left(\frac{-U_{+}}{X_{+}}\right)^{2k}\right)\left(\frac{V_{+}}{U_{+}} \right)^{k-1} \frac{4eE_{+}^2}{U_{+}^2},  
\end{split}
\end{equation}
where the first and third lines of Eq.~(\ref{dV_second_WH}) are used to obtain the second equality, while Eqs.~(\ref{slope_kruskal_out}) and (\ref{slope_kruskal_Horizon_WH}) are used to have the final expression. 
In this form, we can clearly see that at the white hole event horizon $V_{+}=0$, the slope is either equal to zero if $k>1$, or divergent if $k<1$. However, from Eq.~(\ref{slope_kruskal_Horizon_WH}) we have seen that in the Kruskal diagram, the slope is always finite at the white hole event horizon except at the bifurcation surface. Therefore, Eq.~(\ref{problem_at_horizon}) shows a new problem of the resulting Penrose diagram as the cost of fixing the discontinuity at the thin shell. That is, a degeneracy of the four-velocities at the event horizon.


\section{The Bounded Radial Geodesics in the Schwarzschild-to-de Sitter Transition}\label{Sec:Sch_to_dS}

Another plausible resolution of the singularity is the Schwarzschild-to-de Sitter transition.  A rudimentary but general consideration of this kind of scenario is also provided by the thin-shell approximation through an effective static spacelike thin shell \cite{Frolov:1988vj, Balbinot:1990zz}. The studies of the corresponding junction conditions and the stability issue of this transition are typically done by using the static coordinates of the de Sitter spacetime, in which the metric is given by 
\begin{equation}\label{dS_static_metric}
ds^2=-f_{dS}(r)dt^2+f_{dS}^{-1}(r)dr^2+r^2d\Omega^2, 
\end{equation}
with 
\begin{equation}
f_{dS}(r)=1-\frac{r^2}{\ell^2}, 
\end{equation}
where $\ell$ is the characteristic length scale of a given de Sitter space. Then, the static spacelike thin shell connecting the Schwarzschild phase to the de Sitter phase has coordinates $r_{-}=r_{+}=b$, where the convention for the subscripts of the $r$-coordinates follows the one mentioned in Fig. \ref{fig:+&-_convention}.

Similarly, in the corresponding Penrose diagram, an implicit discontinuity generally exists at the thin shell by the simple cut-and-paste procedure; therefore,  additional transformation is required to fix it. This additional transformation also generates a similar coordinate singularity at the event horizons due to the same reason as the connected Penrose diagram of the generalized black-to-white hole bounce \cite{Lin:2023ztq}. In the following, we first review the KS-like coordinates for de Sitter space, and then derive the trajectory equations for a radial geodesic in this coordinate system by the method given in Sec. \ref{Sec: Bounded_Radial_Geodesics_Schwarzschild}. We next quickly mention the transformations required to generate a Penrose diagram of the Schwarzschild-to-de Sitter transition, and then discuss the corresponding choice of constants of integration similar to the procedure given in Sec.~\ref{subsec: bounded_radial_geodesic_BWH}. In the last part of this section, we show, in both the coordinate-independent and coordinate-dependent ways, that the smoothly-crossing of a radial geodesic through the thin shell leads to the same energy-shifting relation. The consistency of those two methods shows that the coordinates related to the connected Penrose diagram form a well-behaved coordinate system covering the entire thin shell with its neighborhood.


\subsection{Radial geodesics in the null KS-like coordinates of de Sitter space}\label{subsec:radial geodesic in dS}

\begin{figure}[h!]
	\centering
	\includegraphics[scale=0.4]{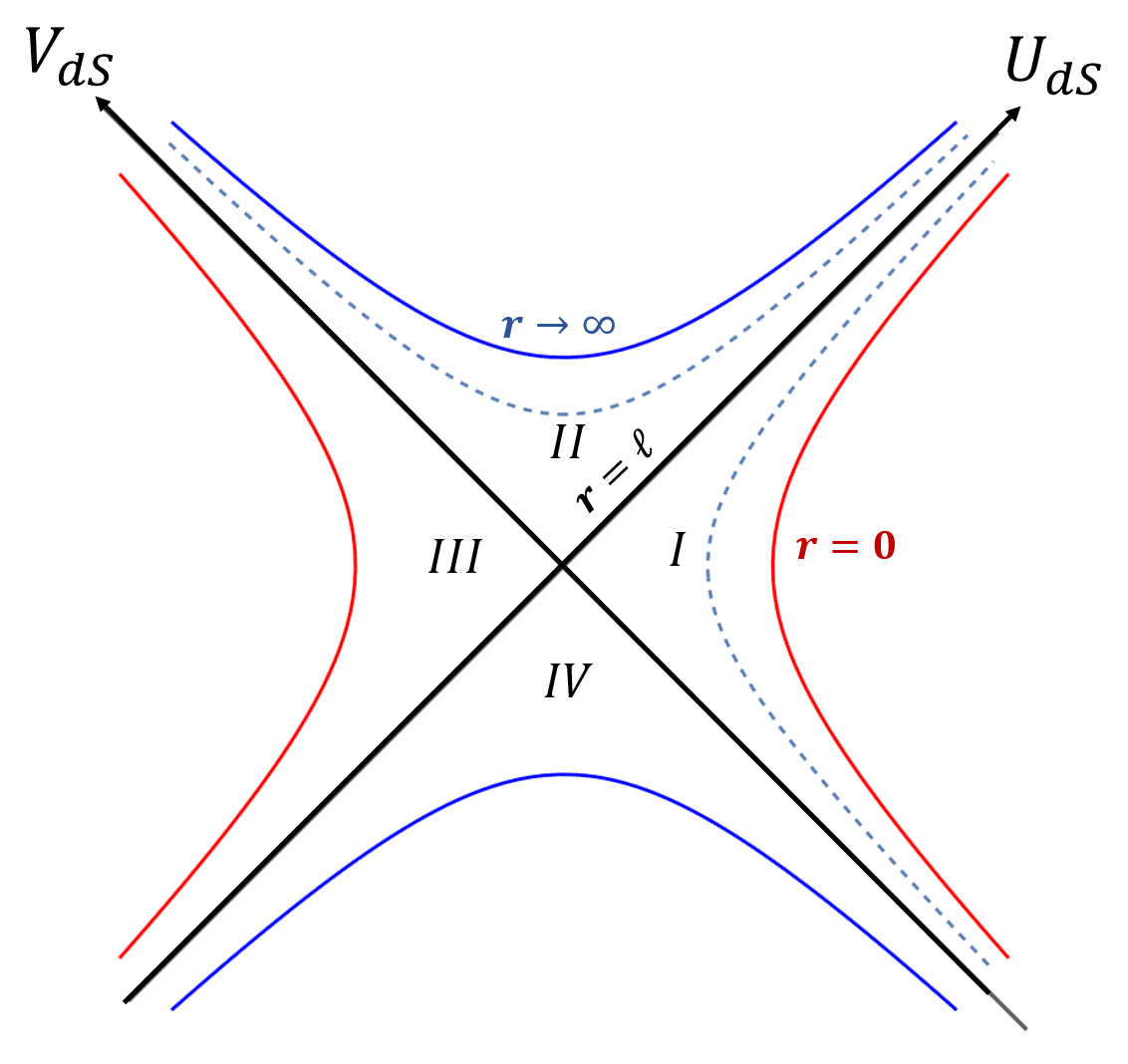}
	\caption{ The Kruskal diagram of a de Sitter space. Notice the difference of the convention of $(U_{dS},V_{dS})$ here and $(V, U)$ in Fig~\ref{fig:Simple_KS}.}  
	\label{fig:deSitter_Kruskal}
\end{figure}

There are many resemblances between the static metric of de Sitter space and the Schwarzschild solution due to the forms of Eqs.~(\ref{dS_static_metric}) and (\ref{Schwarzschild_metric}). Firstly, the static metric (\ref{dS_static_metric}) covers only the region with $r<\ell$, while an event horizon exists at $r=\ell$ for the observer at rest at $r=0$. Meanwhile, this metric also works in the region with $r>\ell$ provided that $r$ becomes the temporal coordinate and $t$ becomes a spatial coordinate. Secondly, 
much similar to the Schwarzschild solution, the coordinate singularity at the de Sitter horizon $r=\ell$ can be removed by introducing the retarded and advanced coordinates: $u_{dS}=t-r^*_{dS}$ and $v_{dS}=t+r^*_{dS}$, where $r^*_{dS}$ is the  
corresponding tortoise coordinate \cite{Spradlin:2001pw, Zee:2013dea, Blau:GRnote}
\begin{equation}\label{tortoise_dS}
r^*_{dS}= \frac{\ell}{2} \log \left|\frac{\ell+r}{\ell-r}\right|. 
\end{equation}
Then, one can introduce the KS-like null coordinates for de Sitter space
\begin{equation}\label{KS_deSitter}
 \left(U_{dS}, V_{dS} \right) \equiv \left(\pm  e^{u_{dS}/\ell} , \pm  e^{-v_{dS}/\ell} \right),  
\end{equation}
where the plus/minus signs are determined by the quadrants considered in the corresponding Kruskal diagram of de Sitter space. See Fig.~\ref{fig:deSitter_Kruskal}. 
By using the KS-like null coordinates from Eq.~(\ref{KS_deSitter}), the de Sitter metric is given as
\begin{equation}
ds^2=\ell^2 \left[\frac{-4}{(1-U_{dS}V_{dS})^2}dU_{dS}dV_{dS} + \frac{(1+U_{dS}V_{dS})^2}{(1-U_{dS}V_{dS})^2}d\Omega^2 \right], 
\end{equation}
where $(U_{dS}, V_{dS})$ satisfy the following relations
\begin{equation}\label{UV_dS}
U_{dS}V_{dS}=\frac{r-\ell}{r+\ell},
\end{equation}
and
\begin{equation}\label{U/V_dS}
\frac{U_{dS}}{V_{dS}}=\pm e^{2t/\ell}.     
\end{equation}

Next, to obtain the trajectory equation of the radial geodesic in the null KS-like coordinates, we start from the four-velocity of a timelike radial geodesic in the static coordinates. Analogous to Eq.~(\ref{radial_geodesic_Deq}), it is given by
\begin{equation}\label{radial_geodesic_Deq_dS}
\mathcal{U}^{\alpha}=\left(\frac{d t}{d \tau}, \frac{d r}{d \tau}, \frac{d \theta}{d \tau}, \frac{d \phi}{d \tau} \right)=\left(\frac{\mathcal{E}}{f_{dS}}, -\epsilon \sqrt{\mathcal{E}^2-f_{dS}}, 0, 0\right), 
\end{equation}
where $\tau$ and $\mathcal{E}$ are the proper time and the energy parameter associated with the geodesic, respectively. Also, we have $\epsilon=1$ for the radially ingoing geodesics, and $\epsilon=-1$ for the radially outgoing geodesics in order to be consistent with the convention used in the Schwarzschild spacetime. 

We now solve Eq.~(\ref{radial_geodesic_Deq_dS}) using the same procedure given in Secs.~\ref{SubSec:radial_infalling_in_Null_KS} and~\ref{SubSec:radial_outgoing_in_Null_KS}. In the following, we also keep the notations and the forms of relations close to those used in the above-mentioned two sections for easier comparison. The $r-$component of Eq.~(\ref{radial_geodesic_Deq_dS}) gives
\begin{equation}\label{dS_radial_geodesics_r-component}
\tau-\tau_0= \int \frac{-\epsilon}{\sqrt{\mathcal{E}^2-f_{dS}}}dr =\frac{-\epsilon \ell}{2} \log \left|\frac{\sqrt{\mathcal{E}^2-f_{dS}}+\frac{r}{\ell}}{\sqrt{\mathcal{E}^2-f_{dS}}-\frac{r}{\ell}}\right| \equiv \epsilon A_{dS}(r), 
\end{equation}
with $f_{dS}=1-r^2/\ell^2$. When $\mathcal{E}<1$, there is a minimal radius the geodesic can reach $r_{min}=\ell \sqrt{1-\mathcal{E}^2}$ from the condition $\sqrt{\mathcal{E}^2-f_{dS}}=0$, which leads to $A_{dS}(r_{min})=0$. While when $\mathcal{E} \geq 1$, the radial geodesic can reach $r=0$, at which we also have $A_{dS}(0)=0$. Thus regardless the value of $\mathcal{E}$, we always have $A_{dS}(r_{min})=0$.

Next, by solving the differential equation 
\begin{equation}\label{Deq_dS_constant_tau_S}
  \frac{1}{\mathcal{E}}\frac{d\tau}{dr}= \frac{dt}{dr}+\frac{\epsilon}{\mathcal{E}}\frac{\sqrt{\mathcal{E}^2-f_{dS}}}{f_{dS}},
\end{equation}
the constant $\tau$ surface in the $(t, r)-$coordinates is given by
\begin{equation}\label{eq_dS_constant_tau_S}
\frac{1}{\mathcal{E}}\tau =t +\epsilon \left(\Theta_{dS}(r)+C_{dS} \right)   
\end{equation}
with
\begin{equation}\label{Theta_dS}
\Theta_{dS}(r) \equiv
 \frac{\ell}{2} \log\left|\frac{\mathcal{E}+(f_{dS}+\frac{r}{\ell}\sqrt{\mathcal{E}^2-f_{dS}})}{\mathcal{E}-(f_{dS}+\frac{r}{\ell}\sqrt{\mathcal{E}^2-f_{dS}})}\right|+ \frac{\ell}{\mathcal{E}} \log \left|\sqrt{\mathcal{E}^2-f_{dS}}-\frac{r}{\ell}\right|,
\end{equation}
and $C_{dS}$ is the constant of integration. 
Similar to $A(r_{min})$, the value of $\Theta_{dS}(r_{min})$ is given by   
\begin{equation}\label{Theta_{dS}(r_{min}_&_0)}
\Theta_{dS}(r_{min})= \frac{\ell}{2}  \log \left|\frac{1+\mathcal{E}}{1-\mathcal{E}}\right| + \frac{\ell}{\mathcal{E}} \log \sqrt{|\mathcal{E}^2-1|}, 
\end{equation}
where $r_{min}=0$ for $\mathcal{E} \geq 1$, and $r_{min}=\ell \sqrt{1-\mathcal{E}^2}$
for $\mathcal{E}<1$. By subtracting the tortoise coordinate (\ref{tortoise_dS}), we can define 
\begin{equation}\label{S(r)_dS}
S_{dS} \equiv \Theta_{dS}-r^{*}_{dS}+C_{dS},
\end{equation}
which is finite at the horizon. 
Using the above relations with Eq.~(\ref{KS_deSitter}), 
we have, for the radially ingoing geodesic $\epsilon=1$,
\begin{equation}\label{V(r)_dS_ingoing}
V_{dS}(r)=-e^{-(A_{dS}+\tau_0)/(\ell \mathcal{E})}e^{S_{dS}/\ell}
\end{equation}
and 
\begin{equation}\label{U(r)_dS_ingoing}
U_{dS}(r)=\frac{\ell-r}{\ell+r}e^{(A_{dS}+\tau_0)/(\ell \mathcal{E})}e^{-S_{dS}/\ell},
\end{equation}
where $A_{dS}$ is defined in  Eq.~(\ref{dS_radial_geodesics_r-component}). While for the radially outgoing geodesic $\epsilon=-1$, we have  
\begin{equation}\label{V(r)_dS_outgoing}
V_{dS}(r)=-\frac{\ell-r}{\ell+r}e^{(A_{dS}-\tau_0)/(\ell \mathcal{E})}e^{-S_{dS}/\ell}
\end{equation}
and 
\begin{equation}\label{U(r)_dS_outgoing}
U_{dS}(r)=e^{-(A_{dS}-\tau_0)/(\ell \mathcal{E})}e^{S_{dS}/\ell}.
\end{equation}
Analogous to the situation in the Schwarzschild spacetime, by using the above four equations with the corresponding constants $\tau_{0}$ and $C_{dS}$, one can plot radial geodesics in the Kruskal diagram of de Sitter space. Then in the next part, we will utilize those equations together with their counterparts of the Schwarzschild spacetime to generate the trajectories of radial geodesics in the connected Penrose diagram of the Schwarzschild-to-de Sitter transition.     


\subsection{Bounded radial geodesics in the Penrose diagram of Schwarzschild-to-de Sitter transition}\label{subsec:connecting_BH_dS}

\begin{figure}[h!]
	\centering
	\includegraphics[scale=0.2]{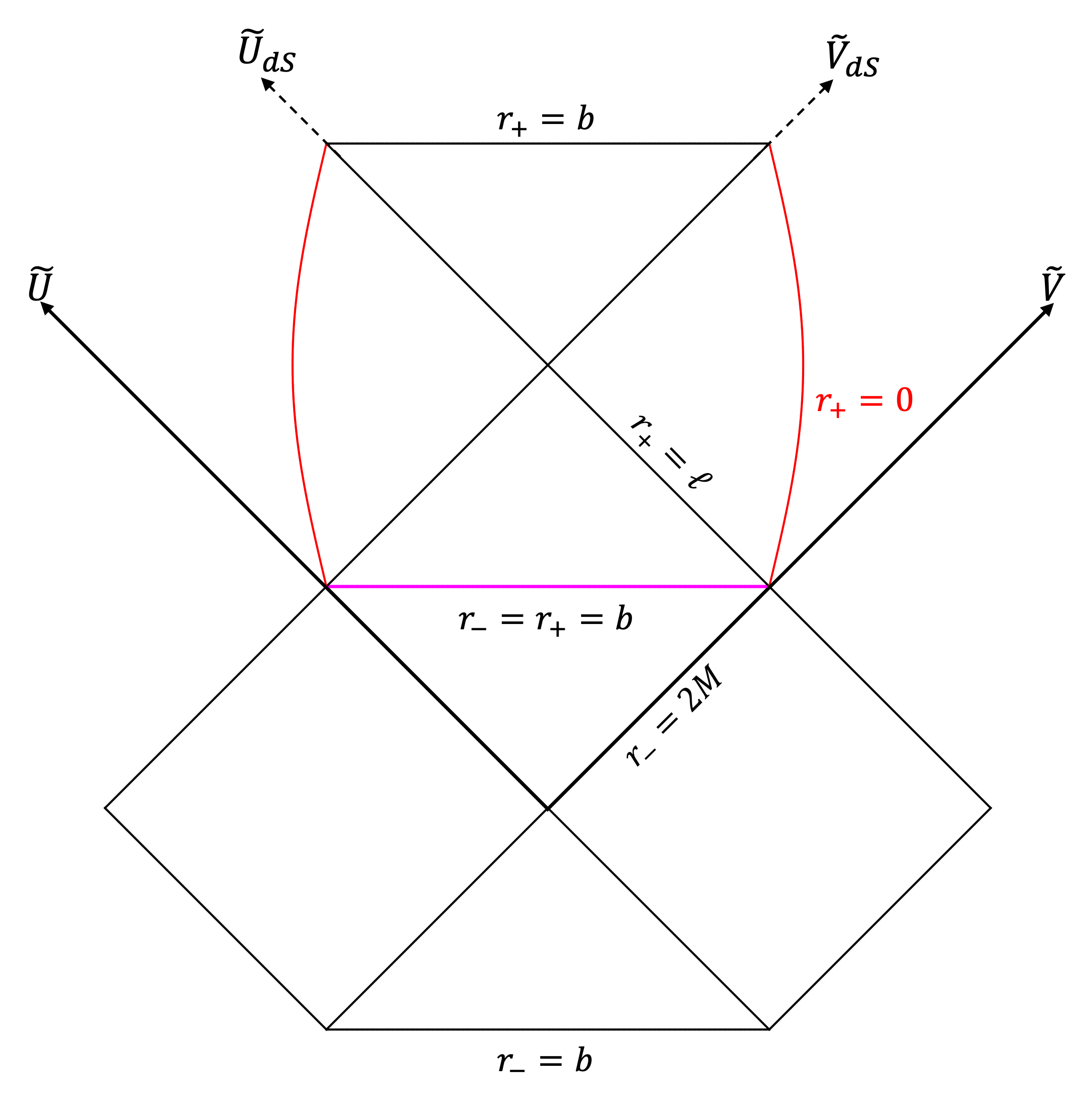}
	\caption{The connected Penrose diagram for the Schwarzschild-to-de Sitter transition with parameters $b=2M \times 7/10$ and $\ell=2M \times 1/6 $.}  
	\label{fig:BH_dS_trans}
\end{figure}

We start by reviewing the transformations required to generate the connected Penrose diagram without illness at the thin shell (located at $r_{-}=r_{+}=b$) introduced in Ref.~\cite{Lin:2023ztq}. With the Schwarzschild part given exactly by the relations labeled by ``$-$'' in Sec.~\ref{subsec:cut_and_paste_procedure_BWH}, we only have to discuss the procedure for the de Sitter phase. 

Firstly, in order to remove the unwanted region $r_{+}>b$ later, we perform a conformal transformation  
\begin{equation}\label{first_conformal_dS}
\left(V'_{dS}, U'_{dS}\right)=\left(\frac{V_{dS}}{X_{dS}}, \frac{U_{dS}}{X_{dS}}\right),
\end{equation}
where $X_{dS}^2 \equiv (b-\ell)/(b+\ell)$. Next, a second transformation is required to fix the discontinuity at the thin shell by enforcing the result of the first junction condition on the resulting Penrose diagram. Notice that this transformation only has to act on one of the two phases. Here we choose to perform the second transformation on the de Sitter phase, and therefore the modified KS-like coordinates (after the two transformations) are given by
\begin{equation}\label{Second_conformal_BH_dS}
\left(V''_{dS}, U''_{dS}\right)=\left(V'_{dS} \left| V'_{dS} \right| ^{k_{dS}-1}, U'_{+} \left| U'_{+} \right| ^{k_{dS}-1} \right),
\end{equation}
where 
\begin{equation}\label{k_dS}
    k_{dS} \equiv \frac{\ell}{4M}\sqrt{\frac{f_{dS}(b)}{f_{BH}(b)}}.
\end{equation}
One can check that the modified KS-like coordinates satisfy the following relation
\begin{equation}\label{second_conformal_dS_t}
\frac{U''_{dS}}{V''_{dS}}=\pm e^{\frac{t_{dS}}{2M}\sqrt{f_{dS}(b)/f_{BH}(b)}}. 
\end{equation}
Then, we can cut out the unwanted region with $r_{+}>b$ after the compactification by the inverse tangent transformation  
\begin{equation}
(\tilde{V}_{dS}, \tilde{U}_{dS})= \left(\tan^{-1}V''_{dS}, \tan^{-1}U''_{dS}\right). 
\end{equation}
Lastly, to paste to the Schwarzschild phase, we use the following identification for the de Sitter phase 
\begin{equation}\label{BH_dS_connection_rule}
(\tilde{V}, \tilde{U}) =
     \left(\tilde{V}_{dS}+\frac{\pi}{2}, \tilde{U}_{dS}+\frac{\pi}{2}\right),    
\end{equation} 
while the identification for the Schwarzschild phase is given by Eq.~(\ref{identifaction_1st_BH}). The corresponding Penrose diagram of this type of spacetime is given in Fig.~\ref{fig:BH_dS_trans}.

\begin{figure}[h!]
	\centering
	\includegraphics[scale=0.3]{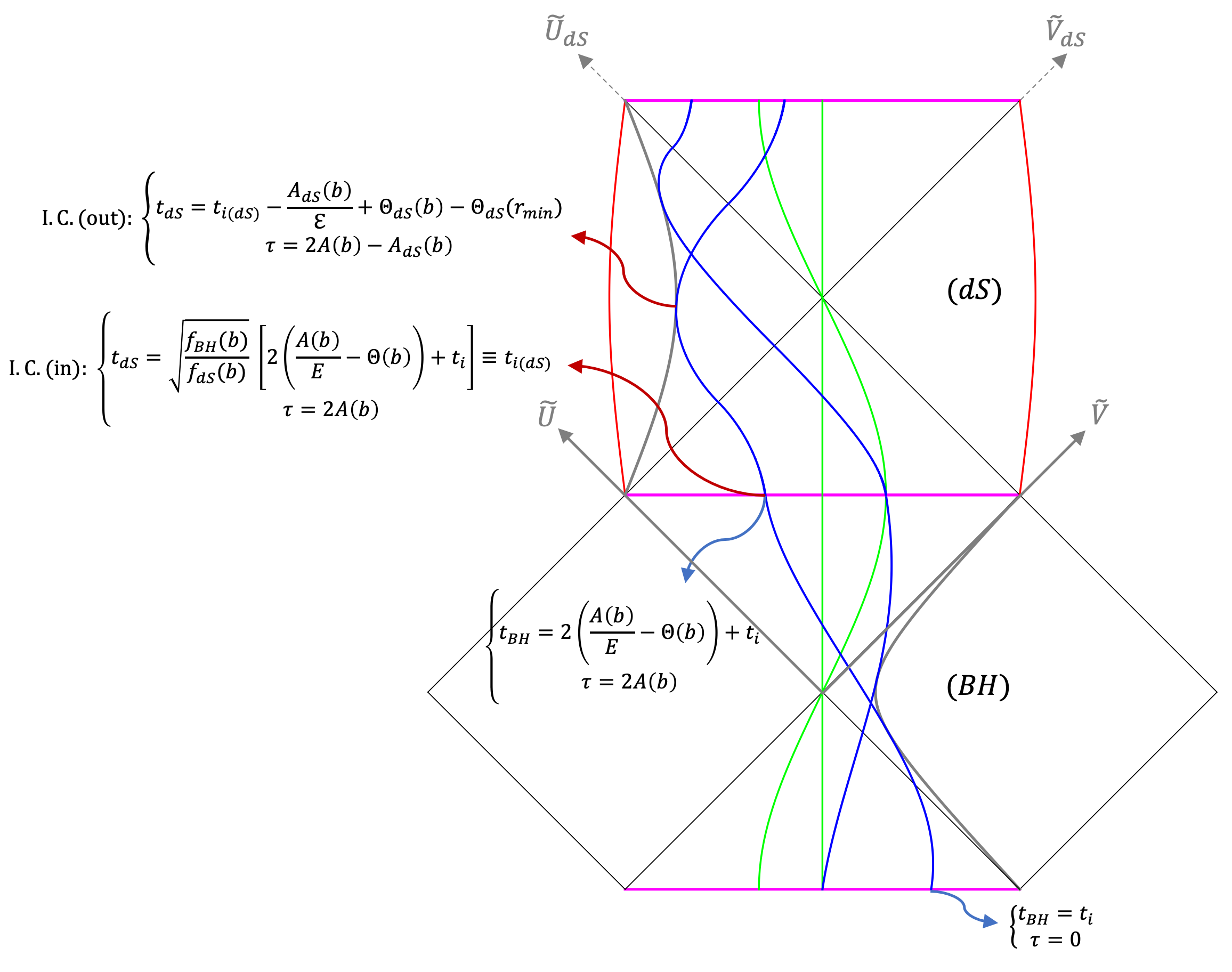}
	\caption{The trajectories of  bounded radial geodesics (blue lines) in the connected Penrose diagram of the Schwarzschild-to-de Sitter transition. The initial conditions for the segments in the de Sitter phase are shown, while those for the segments in the Schwarzschild phase are the same as $I.C. 1$ and $I.C. 2$ given in Fig.~\ref{fig:BWH_geodesics}. Due to the parameters used ($b=2M \times 7/10$ and $\ell=2M \times 1/6 $), the second transformation acting on the de Sitter phase in this diagram makes timelike geodesics become perpendicular to the event horizons in that region, and thus, the coordinate singularity at those event horizons is of the squeezing type. See Ref.~\cite{Lin:2023ztq} for the discussion.}  
	\label{fig:BH_dS_geodesics}
\end{figure}

Now, to construct the trajectory of a radial geodesic in the resulting Penrose diagram of the Schwarzschild-to-de Sitter transition, one applies the transformations, Eqs.~(\ref{first_conformal_dS}) and (\ref{Second_conformal_BH_dS}) to the ingoing and outgoing trajectories Eqs.~(\ref{V(r)_dS_ingoing}) (\ref{U(r)_dS_ingoing}), (\ref{V(r)_dS_outgoing}) and (\ref{U(r)_dS_outgoing}) with the corresponding constants of integration for each segment chosen as follows.  

Again, we choose that all of the geodesics start from the bottom of the diagram, \textit{i.e.} $r_{-}=b$ of the white hole, such that $t_{-}(r_{-}=b)=t_{i}$ and  $\tau(r_{-}=b)=0$. Then for a radial geodesic starting from $t_{i}$, one can show that it crosses the thin shell connecting the black hole to the de Sitter phase at position (in terms of the coordinates of the Schwarzschild side):
\begin{equation}
t_{BH}(r_{-}=b)=2\left(\frac{1}{E}A(b)-\Theta(b)\right)+t_{i},  
\end{equation}
with the elapsed proper time 
\begin{equation}
\tau(r_{-}=b)=2A(b).
\end{equation}
Then the two constants of integration related to this geodesic during the ingoing period in the de Sitter phase, Eqs.~(\ref{V(r)_dS_ingoing}) and (\ref{U(r)_dS_ingoing}), are determined by the conditions 
\begin{equation}
\tau(r_{+}=b)=2A(b),
\end{equation}
and
\begin{equation}\label{t_two_side_relation_dS}
t_{(dS)}(r_{+}=b)=\sqrt{\frac{f_{BH}(b)}{f_{dS}(b)}}t_{BH}(b). 
\end{equation}
Using Eqs.~(\ref{dS_radial_geodesics_r-component}) and (\ref{eq_dS_constant_tau_S}) with the above two conditions, we have the two constants of integration $\tau_{in(dS)}$ and $C_{in(dS)}$ as
\begin{equation}
\tau_{in(dS)}=\tau(b)-A_{dS}(b)=2A(b)-A_{dS}(b)
\end{equation}
and
\begin{equation}
C_{in(dS)}=\frac{1}{\mathcal{E}}\tau(b)-t_{i(dS)}-\Theta_{dS}(b)=\sqrt{\frac{f_{BH}(b)}{f_{dS}(b)}}\left(2\Theta(b)-t_{i}\right)-\Theta_{dS}(b),
\end{equation}
respectively. Also notice that to have the energy shifting relation 
\begin{equation}\label{energy_shift_dS}
\mathcal{E}=\sqrt{\frac{f_{dS}(b)}{f_{BH}(b)}}E
\end{equation}
is used. Lastly, the two constants of integration related to the outgoing period in the de Sitter phase, Eqs.~(\ref{V(r)_dS_outgoing}) and (\ref{U(r)_dS_outgoing}), are determined by the continuity of $(t_{-}, \tau)$ at $r_{+}=r_{min}$. One can show that those two constants are given by  
\begin{equation}
\tau_{out(dS)}=2A(b)-A_{dS}(b)=\tau_{in(dS)},
\end{equation}
and
\begin{equation}
C_{out(dS)}=-C_{in(dS)}-2\Theta_{dS}(r_{min}).
\end{equation}
Using this method, the trajectories of the radial geodesics in the resulting Penrose diagram without illness around the thin shell can be plotted as shown in Fig.~\ref{fig:BH_dS_geodesics}. In the next part, we demonstrate that this resulting Penrose diagram is free from illness at the thin shell by the same argument given in Sec.~\ref{Subsec:derive_energy_shift_BWH}.    


\subsection{Smoothly crossing the thin shell in the Schwarzschild-to-de Sitter transition}\label{subSec:smoothly_crossing_BHtodS}

We first use the coordinate-independent method introduced in Ref. \cite{Hong:2022thd} to show that to have a radial geodesic cross the spacelike thin shell smoothly in the Schwarzschild-to-de Sitter transition scenario, the energy parameters associated with it during the Schwarzschild phase $E$ and the de Sitter phase $\mathcal{E}$ must satisfy the energy shift relation, Eq.~(\ref{energy_shift_dS}), 
\begin{equation}\label{energy_shift_dS_appendix}
\mathcal{E}=\sqrt{\frac{f_{dS}(b)}{f_{BH}(b)}}E,
\end{equation}
where $f_{dS}(r_{+})=1-(r_{+}/\ell)^2$ and $f_{BH}(r_{-})=1-2M/r_{-}$, while $r_{-}=r_{+}=b$ is the ``location'' of the transition surface. After that, in the second part, we show that to have the trajectory of a radial geodesic smoothly crossing the thin shell in the connected Penrose diagram of such a scenario, the same condition Eq.~(\ref{energy_shift_dS_appendix}) must be satisfied. That is, in the second part, we demonstrate this energy shift relation in a coordinate-dependent way. This result then reversely shows the coordinates of the connected Penrose diagram serve as a well-behaved coordinate chart covering the entire thin shell and the spacetime region around it.    

Firstly, the strategy used in the coordinate-independent method is based on the fact that for the Schwarzschild solution, the group of radial geodesics with $E=0$ can exist only inside the event horizon, and their trajectories are orthogonal to the constant $r$ hypersurface \cite{Hong:2022thd}. This special group of radial geodesics has its equivalent in the de Sitter space, which is the group of radial geodesic with $\mathcal{E}=0$ staying only outside of the de Sitter horizon with respect to an observer rest at the north or south pole of the static coordinates (\ref{dS_static_metric}). Then in the Schwarzschild-to-de Sitter transition scenario, a radial geodesic with $E=0$ in the Schwarzschild phase must be a radial geodesic with $\mathcal{E}=0$ in the de Sitter phase after crossing the shell. 

Thus, the smoothly crossing condition of a general radial geodesic can be formulated as the continuity of the inner product of the four-velocity of it $\mathcal{U}_{BH/dS}^{\alpha}$ and the four-velocity of the special radial geodesic mentioned above   $\mathcal{V}_{BH/dS}^{\alpha}$ at the thin shell as 
\begin{equation}\label{Gamma_matching_dS}
\gamma_{BH}(b)=\gamma_{dS}(b),
\end{equation}
where 
\begin{equation}\label{Gamma_limit_Def_BH}
\gamma_{BH}(b) \equiv \lim_{r_{-} \to b} -g_{BH \alpha \beta}\mathcal{U}_{BH}^{\alpha}\mathcal{V}_{BH}^{\beta}, 
\end{equation}
and 
\begin{equation}\label{Gamma_limit_Def_dS}
\gamma_{dS}(b) \equiv \lim_{r_{+} \to b} -g_{dS \alpha \beta}\mathcal{U}_{dS}^{\alpha}\mathcal{V}_{dS}^{\beta}.
\end{equation}
One then can show that to have the relation (\ref{Gamma_matching_dS}), the energy parameters related to $\mathcal{U}_{BH/dS}^{\alpha}$ must satisfy the energy shift relation (\ref{energy_shift_dS_appendix}). 

Next, we show that this energy shift relation (\ref{energy_shift_dS_appendix}) can also be derived in a coordinate-dependent way by using the coordinates of the connected Penrose diagram, \textit{i.e.} the smoothness of the trajectory of a radial geodesic in the connected Penrose diagram at the transition surface
\begin{equation}\label{Smoothly_crossing_condition_dS}
\frac{d\tilde{U}}{d\tilde{V}} \Big|_{r_{-} \to b}=\frac{d\tilde{U}}{d\tilde{V}} \Big|_{r_{+} \to b}.
\end{equation}
The derivation is much similar to the procedure we have done in Sec \ref{Subsec:derive_energy_shift_BWH}, and we use a similar notation by using the quantity $q \equiv 1/\mathcal{E}^2$ in the following. 

We start by expressing the trajectory of an infalling radial geodesic in the null KS coordinates of de Sitter space as 
\begin{equation}\label{V_U_dS_exp_form}
(V_{dS}, U_{dS})=\left(-e^{g_{dS}(r_{+})}, -e^{h_{dS}(r_{+})} \right),
\end{equation}
where $g_{dS}(r_{+})$ and $h_{dS}(r_{+})$ are given by 
\begin{equation}\label{g_{dS}(r)}
g_{dS}(r_{+})=\frac{1}{\ell}\left[-\sqrt{q}(A_{dS}(r_{+})-\tau_{in(dS)})+\Theta_{dS}(r_{+})-r_{dS}^{*}(r_{+})+C_{in(dS)}\right], 
\end{equation}
and 
\begin{equation}\label{h_{dS}(r)}
h_{dS}(r_{+})=\log \left(\frac{r_{+}-\ell}{r_{+}+\ell}\right)-g_{dS}(r_{+}),
\end{equation}
respectively. We also suppress the subscript $+$ and simply use $r$ in the following. Then, by using Eq.~(\ref{V_U_dS_exp_form}), the trajectory of an infalling radial geodesic in the de Sitter phase on the resulting Penrose diagram is given by 
\begin{equation}
(\tilde{V}, \tilde{U})=\left(\tilde{V}_{dS}+\frac{\pi}{2}, \tilde{V}_{dS}+\frac{\pi}{2} \right)= \left(\tan^{-1}[X_{dS}^{k_{dS}}e^{-k_{dS}g_{dS}}], \tan^{-1}[X_{dS}^{k_{dS}}e^{-k_{dS}h_{dS}}] \right), 
\end{equation}
where $k_{dS} = \frac{\ell}{4M}\sqrt{f_{dS}(b)/f_{BH}(b)}$.  By using the following differential relations
\begin{equation}\label{dS_differential relations_A_Theta_r*}
\begin{split}
\frac{dA_{dS}(r)}{dr}&=\frac{-\sqrt{q}}{\sqrt{1-qf_{dS}}},   \\
\frac{d\Theta_{dS}(r)}{dr}&= \frac{\sqrt{1-qf_{dS}}}{f_{dS}}, \\
\frac{dr_{dS}^{*}(r)}{dr}&=\frac{1}{f_{dS}},
\end{split}
\end{equation}
we then have, 
\begin{equation}\label{dg_{dS}/dr}
\frac{dg_{dS}}{dr}= \frac{-1}{\ell}\left(\frac{\sqrt{1-qf_{dS}}-1}{f_{dS}\sqrt{1-qf_{dS}}} \right),
\end{equation}
and 
\begin{equation}\label{df_{dS}/dr}
\frac{dh_{dS}}{dr}=\frac{-1}{\ell}\left(\frac{\sqrt{1-qf_{dS}}+1}{f_{dS}\sqrt{1-qf_{dS}}} \right).
\end{equation}

With a similar calculation, we have, from the side of the de Sitter phase 
\begin{equation}\label{dU/dV_dS}
\begin{split}
\frac{d\tilde{U}}{d\tilde{V}}=\frac{d\tilde{U}/dr}{d\tilde{V}/dr}
=&\frac{1+X_{dS}^{2k_{dS}}e^{-2k_{dS}g_{dS}}}{1+X_{dS}^{2k_{dS}}e^{-2k_{dS}h_{dS}}} \times e^{k_{dS}(g_{dS}-h_{dS})} \times \frac{dh_{dS}/dr}{dg_{dS}/dr} \\
=&\frac{1+\tan^{2}\tilde{V}}{1+\tan^{2}\tilde{U}} \times \frac{V''_{dS}}{U''_{dS}} \times \frac{\sqrt{1-qf_{dS}}+1}{\sqrt{1-qf_{dS}}-1}. 
\end{split}
\end{equation}
Notice that the final form of Eq.~(\ref{dU/dV_dS}) is similar to that of Eq.~(\ref{dU/dV_WH}) besides the factor $V''_{dS}/U''_{dS}$, which is inverse to its counterpart $U''_{+}/V''_{+}$ in Eq.~(\ref{dU/dV_WH}). One can see that this difference is due to the convention used by comparing Eq.~(\ref{U/V_dS}) to Eq.~(\ref{Kruskal_U/V}). Substituting Eqs.~(\ref{dU/dV_BH}) and (\ref{dU/dV_dS}) into Eq.~(\ref{Smoothly_crossing_condition}), together with the relations (\ref{U/V_mini}), (\ref{second_conformal_dS_t}) and (\ref{t_two_side_relation_dS}), we arrive the energy shift relation Eq.~(\ref{energy_shift_dS_appendix})
\begin{equation*}
\frac{f_{BH}(b)}{E^2}=\frac{f_{dS}(b)}{\mathcal{E}^2},
\end{equation*}
where $p_{-}=1/E^2$ and $q=1/\mathcal{E}^2$ are used. 
Then, the consistency of the two
methods shows that the coordinates of the connected Penrose diagram $(\tilde{V}, \tilde{U})$ form a well-behaved coordinate system that covers the entire thin shell with its neighborhood. 


\section{Conclusions}\label{Sec:conclusion}

In this article, we study the trajectories of the bounded radial geodesics in the Penrose diagram for the spacetime constructed via a static spacelike thin shell by using two examples: the generalized black-to-white hole bounce and the Schwarzschild-to-de Sitter transition. In Ref.~\cite{Lin:2023ztq}, it was found that to construct the Penrose diagram of this type of spacetime, two transformations are required in general. The first transformation makes the shell have the same shape in the two coordinate charts that we would like to cut and paste, while the second transformation enforces the first junction condition in the resulting Penrose diagram. In the work presented here, we gave the detailed construction of the bounded radial geodesics trajectories in the resulting Penrose diagrams, and showed that these trajectories indeed cross the thin shell smoothly in the two examples, \textit{e.g.} Fig.~\ref{fig:BWH_geodesics} and Fig.~\ref{fig:BH_dS_geodesics}.\footnote{Although we only demonstrate how to construct the trajectory equation of a bounded radial geodesic crossing one single thin shell,   
it's also not difficult to repeat the procedure introduced to as many cycles as one wishes, \textit{i.e.} phases $\mathcal{M}_{i}$ with $i=1, 2, 3...$, separated by thin shells located at $r_{i}=b$.} Reversely, this is a demonstration that the coordinates of the resulting Penrose diagram form a well-behaved coordinate chart covering the entire thin shell and its neighborhood. However, as pointed out in Ref.~\cite{Lin:2023ztq}, the second transformation enforcing the first junction condition also unavoidably introduces a new type of coordinate singularity back to the event horizons. This new type of singularity causes distortion of the trajectories of timelike radial geodesics at the event horizons under the second transformation as shown in Figs.~\ref{fig:BWH_geodesics}, ~\ref{fig:second_trans_BHtoWH} and ~\ref{fig:BH_dS_geodesics}. We further use the generalized black-to-white hole bounce as an example to analytically show the effect of this coordinate singularity on these trajectories, Eq.~(\ref{problem_at_horizon}). 

Due to this singularity, a global conformal coordinate chart for the spacetime connected via a static spacelike thin shell in general does not exist except for some special cases. And thus, the ``final'' Penrose diagram should be constructed through two coordinate charts with an overlapping region covering the thin shell. See Fig.~10 in Ref.~\cite{Lin:2023ztq}. Nevertheless, as shown in Fig.~\ref{fig:BWH_geodesics} and Fig.~\ref{fig:BH_dS_geodesics}, the trajectory of a bounded radial geodesic is still continuous at the event horizons of the upper half of the diagram, and where it crosses the thin shell can be determined unambiguously. Then, at the cost of surrendering the conformal property, one can construct a global coordinate chart for such a spacetime by erasing the effect of the second transformation at the event horizon,  which we would like to address in future work. Lastly, we have discussed spacelike thin shells in this work, but a similar analysis might be applied to timelike shells. Usually, the timelike shells are essential when considering dynamical situations, e.g., collapsing, bounding, or oscillating cases. For all cases, the detailed geodesic or metric description that crossover the shell will be very interesting and with diverse applications, not only classical but also quantum. We also leave this topic for future work.

\newpage

\section*{Acknowledgment}
The authors would like to thank Dejan Stojkovic for useful suggestions. W.L. is supported by the National Research Foundation of Korea (NRF) grant funded by the Korean government (MSIT) (2021R1A4A5031460). DY is supported by the National Research Foundation of Korea (Grant No.: 2021R1C1C1008622, 2021R1A4A5031460). W.L. acknowledges hospitality by University at Buffalo (his alma mater) during the completion of some of this work. W.L. is also grateful to the participants of the $100+7$ GR and Beyond: Inflation workshop by Jeju National University for useful comments and discussions.

\appendix

\section{A review of the coordinate systems formulated by the timelike radial infalling geodesics }\label{Sec:Review}

Before solving Eq.~(\ref{constant_tau_surface_Deq}), one should notice that the different forms of the Schwarzschild metric in the above-mentioned coordinate systems can be obtained by utilizing the fact that Eq.~(\ref{constant_tau_surface_Deq}) is also a total differential relation. That is, one first rewrite it as (with $\epsilon=1$)
\begin{equation}\label{total_Deq_tau_r_t}
d t= \frac{1}{E}\left(d \tau -\frac{\sqrt{E^2-f}}{f}d r \right), 
\end{equation}
and then substitute it into Eq.~(\ref{Schwarzschild_metric}). With some algebraic manipulations, one obtains the Schwarzschild metric in the Gautreau-Hoffmann coordinates as 
\begin{equation}\label{GH_metric}
d s^2= -d \tau^2 +\frac{1}{E^2} \left(\sqrt{E^2-f}d \tau+d r \right)^2 +r^2 d \Omega^2,
\end{equation}
which, by setting $E=1$, reduces to the Painlev{\'e}-Gullstand form
\begin{equation}\label{PG_metric}
d s^2= -d \tau^2 + \left(\sqrt{2M/r}d \tau+d r \right)^2 +r^2 d \Omega^2. 
\end{equation}
Next, one can further diagonalize the metric by choosing a group of geodesics as the new spatial coordinate lines to replace $r$.\footnote{A discussion of applying this method to diagonalize the Painlev{\'e}-Gullstand metric and the properties of those comoving metrics can be found in Ref. \cite{Blau:GRnote}.} That is, since we have used $(\tau, r)$ as the new coordinates, the integral constant $\tau_0$ in Eq.~(\ref{tau_r_relation}) can be viewed as labels of different members in a given family of geodesics specified by $E$. Thus, by promoting the label of each geodesic $\tau_0$ to a new variable $\rho$, we rewrite Eq.~(\ref{tau_r_relation}) into the form
\begin{equation}\label{rho_tau_r_relation}
\tau-\rho=A(r),
\end{equation}
which allows us to further write down the following differential relation
\begin{equation}\label{dr_in_dtau_drho}
\begin{split}
d r= &\frac{\partial r}{\partial \tau} d \tau +\frac{\partial r}{\partial \rho} d \rho \\
= & \frac{d r}{d \tau}(d\tau -d\rho) \\
= &-\sqrt{E^2-f}(d\tau -d\rho), 
\end{split}
\end{equation}
where Eq.~(\ref{radial_geodesic_Deq}) is used to obtain the last line. Using Eq.~(\ref{dr_in_dtau_drho}) to replace the $dr$ in Eq.~(\ref{GH_metric}), one obtain the diagonalized Gautreau-Hoffmann metric
\begin{equation}\label{diag_GH_metric}
ds^2=-d \tau^2 +\frac{E^2-f}{E^2}d \rho^2 +r^2 d \Omega^2,
\end{equation}
which reduces to the Lema{\^i}tre metric when $E=1$: 
\begin{equation}\label{Lemaitre_metric}
ds^2=-d \tau^2 +\frac{2M}{r(\tau, \rho)}d \rho^2 +r^2 d \Omega^2. 
\end{equation}
In Appendix~\ref{Sec:KS_as_limit_of_GH}, we show that Kantowski-Sachs spacetime can be treated as a limit of the diagonalized Gautreau-Hoffmann metric. 

\section{Kantowski-Sachs spacetime as a limit of the diagonalized Gautreau-Hoffmann metric}\label{Sec:KS_as_limit_of_GH}

Due to the exchanging characters of $t$ and $r$ of the Schwarzschild metric (\ref{Schwarzschild_metric}) inside the event horizon, one can rewrite the metric  by the coordinates redefinition, $\bar{r} \equiv t$ and $ \bar{t} \equiv r$, to have 
the interior metric as
\begin{equation}\label{Normal_Schwarzschild_interior_metric}
d s^2=-\left(\frac{2M}{\bar{t}}-1 \right)^{-1}d \bar{t}^2+ \left(\frac{2M}{\bar{t}}-1 \right)d \bar{r}^2+\bar{t}^2d\Omega^2, 
\end{equation}
in which the temporal coordinate $\bar{t}$ runs backward from $\bar{t}=2M$ to $\bar{t}=0$. This is equivalent to identifying the interior of a black hole as a special type of cosmological model: the Kantowski-Sachs spacetime. For a review of this viewpoint, see Ref. \cite{Doran:2006dq}. One can further rewrite the metric into the comoving form: 
\begin{equation}\label{Reg_Schwarzschild_interior_metric}
d s^2=-d \tilde{t}^2+\mathcal{A}^2(\tilde{t})d \tilde{r}^2+\mathcal{B}^2(\tilde{t})d\Omega^2, 
\end{equation}
where 
\begin{equation}\label{KS_comving_A}
\mathcal{A}(\tilde{t})=\tan \eta(\tilde{t}), 
\end{equation}
and
\begin{equation}\label{KS_comving_B}
\mathcal{B}(\tilde{t})=2M \cos^2 \eta \left(\tilde{t} \right), 
\end{equation}
with $\eta(\tilde{t})$ defined implicitly by
\begin{equation}
\tilde{t}=2M(\eta+\sin \eta \cos \eta).
\end{equation}
In this particular form, the new temporal coordinate $\tilde{t}$ is the proper time measured by those observers at rest ($d\tilde{r}=d\phi=d\theta=0$) in the Kantowski-Sachs spacetime. Then the trajectories of those observers in this spacetime define the corresponding comoving reference frame similar to that in a Robertson-Walker spacetime.  In the following, we show that this form of metric corresponds to the $E \rightarrow 0$ limit of the diagonalized Gautreau-Hoffmann metric (\ref{diag_GH_metric}) upon a scaling of the ``spatial'' coordinate $\rho$.

\begin{figure}[h!]
	\centering
	\includegraphics[scale=0.18]{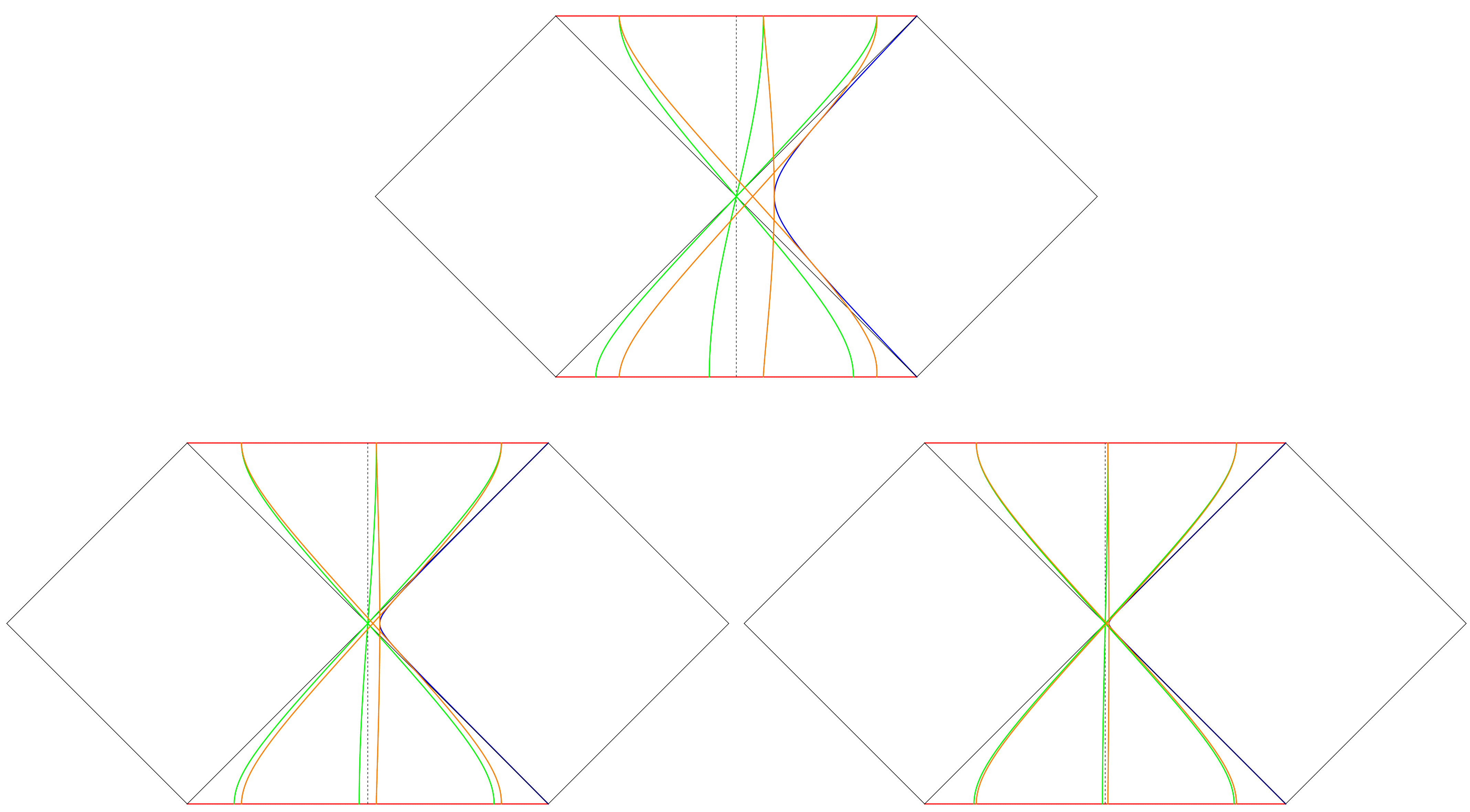}
	\caption{Pictures show that the geodesics (orange lines) approach the trajectories of the comoving observers (green lines) in the Kantowski-Sachs spacetime when $E \to 0$, where $E=1-2M/R$.}  
	\label{fig:Approaching_Kantowski_Sachs}
\end{figure}
Firstly, by comparing Eq.~(\ref{Schwarzschild_metric}) with Eq.~(\ref{Reg_Schwarzschild_interior_metric}) and noticing that angular parts of the metrics must be the same, we have the following relation
\begin{equation}\label{cos_eta_r_relation}
\cos \eta(\tilde{t})=\sqrt{\frac{r}{2M}}, 
\end{equation}
which leads to 
\begin{equation}\label{tan_eta_r_relation}
\tan \eta(\tilde{t})=\sqrt{\frac{2M}{r}-1}=\sqrt{-f}. 
\end{equation}
On the other hand, the $E \rightarrow 0$ limit of the diagonalized Gautreau-Hoffmann metric (\ref{diag_GH_metric}) involves a divergent component:
\begin{equation}\label{limit_divergent_Dia_GH}
\lim_{E\to 0} \sqrt{\frac{E^2-f}{E^2}} d\rho, 
\end{equation}
which can be adsorbed into the new coordinate by a rescaling $\tilde{\rho}=\rho/E$ with the corresponding metric given as
\begin{equation}\label{dia_GH_in_rho'}
ds^2=-d \tau^2 +(E^2-f)d \tilde{\rho}^2 +r^2 d \Omega^2. 
\end{equation}
With Eqs.~(\ref{cos_eta_r_relation}) and (\ref{tan_eta_r_relation}), we see that this form of metric (\ref{dia_GH_in_rho'}) is indeed equivalent to the comoving form of the Kantowski-Sachs metric (\ref{Reg_Schwarzschild_interior_metric}) when $E \to 0$. A plotting of the geodesics with different $E$ values approaching to the trajectories of comoving observers in the Kantowski-Sachs spacetime is shown in Fig. \ref{fig:Approaching_Kantowski_Sachs}.  


\section{The four-velocity of a radial geodesic in the null Kruskal-Szekeres coordinates}\label{Sec: four_velocity_in_KS}
We first work out the result Eq.~(\ref{four_velocity_at_horizon}) from Eqs.~(\ref{four-velocity_V(tau_r)}) and (\ref{four-velocity_U(tau_r)}). 
For convenience, we write Eqs.~(\ref{four-velocity_V(tau_r)}) and (\ref{four-velocity_U(tau_r)}) down again here: 
\begin{equation}\label{four-velocity_V(tau_r)_App}
\frac{d V}{d \tau}=\frac{\partial V}{\partial \tau}+\frac{\partial V}{\partial r}\frac{d r}{d \tau}=\frac{V}{4Mf} \left(E-\sqrt{E^2-f} \right)=\frac{VE}{4Mf} \left(1-\sqrt{1-\frac{f}{E^2}} \right), 
\end{equation}
and
\begin{equation}\label{four-velocity_U(tau_r)_App}
\frac{d U}{d \tau}=\frac{\partial U}{\partial \tau}+\frac{\partial U}{\partial r}\frac{d r}{d \tau}=\frac{-U}{4Mf}\left(E+\sqrt{E^2-f}\right)=\frac{-UE}{4Mf}\left(1+\sqrt{1-\frac{f}{E^2}}\right). 
\end{equation}
We can derive the result around the event horizon by using $r=2M+\delta$, where $\delta \rightarrow 0^{\pm}$ represents approaching the event horizon from outside ($0^+$) and inside ($0^-$) of the black hole respectively. 
\begin{equation}
 f=1-\frac{2M}{r}=1-\frac{2M}{2M+\delta} \rightarrow \frac{\delta}{2M},  
\end{equation}
which gives 
\begin{equation}
\sqrt{1-\frac{f}{E^2}} \rightarrow 1-\frac{1}{2}\frac{f}{E^2} \rightarrow 1-\frac{\delta}{4ME^2}. 
\end{equation}
On the other hand, by using Eq.~(\ref{Kruskal_UV(r)}) we have 
\begin{equation}\label{U_horizon_limit}
U=\left(1-\frac{r}{2M}\right)e^{r/2M}\frac{1}{V}=\left(1-\frac{2M+\delta}{2M}\right)e^{(2M+\delta)/2M}\frac{1}{V} \rightarrow \left(\frac{-\delta}{2M}\right)\frac{e}{V}
\end{equation}
Substituting the above results into Eqs.~(\ref{four-velocity_V(tau_r)_App}) and (\ref{four-velocity_U(tau_r)_App}), we then have the four-velocity of an infalling radial geodesic at the event horizon ($U=0$):  
\begin{equation}\label{four_velocity_at_horizon_App}
\left( \frac{d V}{d \tau}, \frac{d U}{d \tau}\right)  \rightarrow \left( \frac{V}{8ME}, \frac{eE}{2MV}\right),
\end{equation}
in which both components are finite unless $V=0$ at the same time.

By using Eqs.~(\ref{V_(tau, r)_WH}) and (\ref{U_(tau, r)_WH}) the components of the four-velocity of an outgoing radial timelike geodesic can be derived by the same method:     
\begin{equation}\label{Outgoing_four-velocity_V(tau_r)_App}
\frac{d V}{d \tau}=\frac{VE}{4Mf}\left(1+\sqrt{1-\frac{f}{E^2}} \right), 
\end{equation}
and
\begin{equation}\label{Outgoing_four-velocity_U(tau_r)_App}
\frac{d U}{d \tau}=\frac{-UE}{4Mf} \left(1-\sqrt{1-\frac{f}{E^2}} \right). 
\end{equation}
Notice that the white hole horizon is given by $V=0$ instead of $U=0$, so we rewrite $V$ this time. Around the white hole event horizon, it is given by 
\begin{equation}\label{V_horizon_limit}
V=\left(1-\frac{r}{2M}\right)e^{r/2M}\frac{1}{U} \rightarrow \left(\frac{-\delta}{2M}\right)\frac{e}{U}. 
\end{equation}
With a similar calculation, we have the four-velocity of an outgoing radial geodesic at the white hole event horizon ($V=0$):  
\begin{equation}\label{Outgoing_four_velocity_at_horizon_App}
\left( \frac{d V}{d \tau}, \frac{d U}{d \tau}\right)  \rightarrow \left(  \frac{-eE}{2MU}, \frac{-U}{8ME}\right),
\end{equation}
which is again finite except at the bifurcation surface. 

\section{The derivation of Eq.~(\ref{dU/dV_WH})}\label{Sec: the derivation}
In the following, we give a detailed step-by-step derivation of $d\tilde{V}/dr$ in Eq.~(\ref{dU/dV_WH}), while the other factor $d\tilde{U}/dr$ can be derived similarly.

Firstly, the $\tilde{V}$ component of the second white hole part is given by   $\tilde{V}=\tilde{U}_{+}+\frac{\pi}{2}$.  Also notice that for the white hole part  $\tilde{U}_{+}<0$. When $U_{+}<0$, 
\begin{equation}
\tilde{U}_{+}=\tan^{-1} U_{+}'' =\tan^{-1}\left[-\left(-\frac{U_{+}}{X_{+}}\right)^{k}\right] \equiv  \tan^{-1}\left(-\alpha \right)=-\tan^{-1} \alpha,  
\end{equation}
where $\alpha \equiv \left(-\frac{U_{+}}{X_{+}}\right)^{k} =-U_{+}'' >0$. So we have
\begin{equation}
\tilde{V}=\tilde{U}_{+}+\frac{\pi}{2}
=-\tan^{-1} \alpha+\frac{\pi}{2} 
= \tan^{-1}\frac{1}{\alpha}, 
\end{equation}
which gives $\tan \tilde{V} = 1/\alpha$. 
\begin{equation}\label{detailed_calculation}
 \begin{split}
 \frac{d\tilde{V}}{dr}&= \frac{d}{dr} \tan^{-1}\frac{1}{\alpha} \\
 &= \frac{-1}{1+\alpha^2} \frac{d \alpha}{dr} \\
 &=  \frac{-1}{1+\left(-\frac{U_{+}}{X_{+}}\right)^{2k}} \frac{d}{dr} \left(-\frac{U_{+}}{X_{+}}\right)^{k}\\
 &= \frac{-1}{1+\left(-\frac{U_{+}}{X_{+}}\right)^{2k}}  \left[ k\left(-\frac{U_{+}}{X_{+}}\right)^{k-1}\left(\frac{-1}{X_{+}}\right)\frac{dU_{+}}{dr}\right] \\
 &= \frac{k}{1+\left(-\frac{U_{+}}{X_{+}}\right)^{2k}}  \frac{1}{X_{+}^k}\left[\left(-U_{+}\right)^{k-1}\frac{dU_{+}}{dr}\right] \\
 &= \frac{k}{1+\left(-\frac{U_{+}}{X_{+}}\right)^{2k}}  \frac{1}{X_{+}^k}\left[-e^{kg_{+}}\frac{dg_{+}}{dr}\right] \\
 &= \frac{-k}{1+\left(-\frac{U_{+}}{X_{+}}\right)^{2k}}  \left(\frac{-U_{+}}{X_{+}}\right)^k\frac{dg_{+}}{dr} \\
\end{split}
\end{equation}
where $\alpha= \left(-U_{+}/X_{+}\right)^{k}$ is used in the third line, and $U_{+}=-\exp{[g_{+}(r)]}$ is used to obtain the last two lines. Now, by using $\alpha= \left(-U_{+}/X_{+}\right)^{k}$ again, we have
\begin{equation}\label{dV_second_WH}
\begin{split}
\frac{d\tilde{V}}{dr}&= \frac{-k}{1+\alpha^{2}}  \alpha \frac{dg_{+}}{dr} \\
 &= \frac{-k}{1+\alpha^{-2}}  \frac{1}{\alpha}\frac{dg_{+}}{dr} \\
  &= \frac{k}{1+\tan^2 \tilde{V}}  \frac{1}{U_{+}''}\frac{dg_{+}}{dr}, \\
\end{split}
\end{equation}
where $\tan \tilde{V}=1/\alpha$ and $\alpha=-U''_{+}$ are used to obtain the final form.

By a similar derivation, we have 
\begin{equation}\label{dU_second_WH}
 \frac{d\tilde{U}}{dr} = \frac{d}{dr} \left(\tilde{V}_{+}+\frac{\pi}{2}\right) 
 = \frac{k}{1+\tan^2\tilde{U}_{+}} \frac{1}{V_{+}''} \frac{dh_{+}}{dr},
\end{equation}
where $V_{+}=-\exp{[h_{+}(r)]}$ is used. From Eqs.~(\ref{dV_second_WH}) and (\ref{dU_second_WH}), we then have Eq.~(\ref{dU/dV_WH}). 


\bibliographystyle{unsrt}
\bibliography{Bibli_BRG}

\end{document}